\documentclass[11pt,onecolumn,twoside]{IEEEtran}

\usepackage{amsmath, amssymb, graphicx, cite}

\newtheorem{thm}{Theorem}

\newtheorem{defin}{Definition}
\newtheorem{prop}{Proposition}
\newtheorem{lemma}{Lemma}

\newcommand{\tr}{{d_{\tt{c}}}}
\newcommand{\tl}{{d_{\tt{v}}}}
\newcommand{\ctov}{\tt{c}\to\tt{v}}
\newcommand{\vtoc}{\tt{v}\to\tt{c}}
\newcommand{\E}[1]{E\left[{#1}\right]}
\newcommand{\etanh}{\operatorname{etanh}}
\newcommand{\opt}{\text{opt}}

\begin{document}
\title{Performance of LDPC Codes Under \\Faulty Iterative Decoding}

\author{Lav~R.~Varshney
\thanks{Manuscript prepared May 2008; revised April 2009, May 2010.  This work was supported in part by a
National Science Foundation Graduate Research Fellowship and was performed in part when the author was
with l'\'{E}cole Polytechnique F\'{e}d\'{e}rale de Lausanne.  The material in this paper was presented
in part at the Information Theory Workshop, Lake Tahoe, CA, September 2007.}
\thanks{L.~R.~Varshney is with the Department of Electrical Engineering and Computer Science, 
the Laboratory for Information and Decision Systems, and the Research Laboratory of Electronics,
Massachusetts Institute of Technology, Cambridge,
MA, 02139 USA (e-mail: lrv@mit.edu).}
}

\maketitle

\begin{abstract}
Departing from traditional communication theory where decoding algorithms are assumed to perform without error,
a system where noise perturbs both computational devices and communication channels is considered here.
This paper studies limits
in processing noisy signals with noisy circuits by investigating
the effect of noise on standard
iterative decoders for low-density parity-check codes.
Concentration of decoding performance around its average
is shown to hold when noise is introduced into message-passing
and local computation.
Density evolution equations 
for simple faulty iterative decoders are derived.  
In one model, computing nonlinear estimation thresholds shows 
that performance degrades smoothly as decoder noise increases, but
arbitrarily small probability of error is not achievable.   
Probability of error may be driven to zero in another system model; 
the decoding threshold again decreases smoothly with decoder noise.
As an application of the methods developed, an achievability result for
reliable memory systems constructed from unreliable components
is provided.
\end{abstract}


\section{Introduction}
\label{sec:intro}

The basic goal in channel coding is
to design encoder-decoder pairs that allow reliable
communication over noisy channels at information rates close 
to capacity \cite{Shannon1948}.  The 
primary obstacle in the quest for practical capacity-achieving codes 
has been decoding complexity \cite{Blahut1983,Calderbank1998,CostelloF2007}.  
Low-density parity-check (LDPC) codes have, however, emerged as a class of codes
that have performance at or near the Shannon 
limit \cite{PfisterSU2005,ChungFRU2001}
and yet are sufficiently structured as to have decoders 
with circuit implementations \cite{LoeligerLHT2001,BlanksbyH2002,ZhangP2003}.

In addition to decoder complexity, decoder reliability may 
also limit practical channel coding.\footnote{One
may also consider the effect of encoder complexity \cite{BazziM2005},
however encoder noise need not be explicitly considered, since
it may be incorporated into channel noise, using the noise combining argument
suggested by Fig.~\ref{fig:combinenoise}.}  In Shannon's 
schematic diagram of a general communication system 
\cite[Fig.~1]{Shannon1948} and in the traditional information  
and communication theories that have developed within the 
confines of that diagram, noise is localized in the 
communication channel.  The decoder is assumed to
operate without error.  Given the possibility of unreliable 
computation on faulty hardware, there is value in studying
error-prone decoding.
In fact Hamming's original development of parity-check codes 
was motivated by applications in computing rather than in 
communication \cite{Hamming1950}.

\begin{figure}
  \centering
  \includegraphics[width=2.7in]{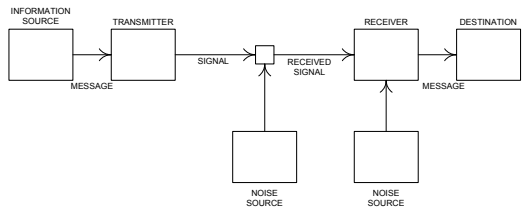}
  \caption{Schematic diagram of an information system that processes unreliable signals with unreliable circuits.}
  \label{fig:fig1_rev}
\end{figure}

The goal of this paper is to investigate 
limits of communication systems with noisy decoders and
has dual motivations.  The first is the eminently practical
motivation of determining how well error control codes work
when decoders are faulty.  The second is the deeper
motivation of determining fundamental limits for processing unreliable
signals with unreliable computational devices, illustrated schematically in Fig.~\ref{fig:fig1_rev}. 
The motivations are intertwined.  As noted by Pierce, ``The down-to-earth problem 
of making a computer work, in fact, becomes tangled with this difficult 
philosophical problem: `What is possible
and what is impossible when unreliable circuits are used to process 
unreliable information?'" \cite{Pierce1965}.

A first step in understanding these issues 
is to analyze a particular class of codes and decoding techniques: 
iterative message-passing decoding algorithms for LDPC codes.  
When the code is represented as a factor graph, algorithm computations occur at nodes 
and algorithm communication is carried out over edges.
Correspondence between the factor graph and the algorithm
is not only a tool for exposition but also the way 
decoders are implemented 
\cite{LoeligerLHT2001,BlanksbyH2002,ZhangP2003}.
In traditional performance analysis, the decoders are assumed 
to work without error.  In this paper, there will be 
transient local computation and message-passing errors, whether the 
decoder is analog or digital.

When the decoder itself is noisy, one might believe that achieving
arbitrarily small probability of error (\emph{Shannon reliability}) is
not possible, but this is indeed possible for certain 
sets of noisy channels and noisy decoders.  This is shown by example.  
For other sets of noisy channels and noisy decoders, Shannon reliability is not
achievable, but error probability tending to extremely small values is achievable.
Small probability of error, $\eta$, is often satisfactory in practice, and so 
$\eta$-reliable performance is also investigated.  Decoding thresholds at
$\eta$-reliability decrease smoothly with increasing decoder noise. 
Communication systems may display graceful degradation with respect to noise levels in the decoder.  

The remainder of the paper is organized as follows.  Section~\ref{sec:background}
reviews motivations and related work.  Section~\ref{sec:notation} formalizes
notation and Section~\ref{sec:tools} gives concentration results that allow the 
density evolution method of analysis, generalizing results in \cite{RichardsonU2001}.  A noisy version of
the Gallager A decoder for processing the output of a binary symmetric channel is analyzed 
in Section~\ref{sec:ex1}, where it is shown that Shannon reliability is unattainable.  In Section~\ref{sec:ex2}, a noisy
decoder for AWGN channels is analyzed.  For this model, the probability of error
may be driven to zero and the decoding threshold degrades smoothly
as a function of decoder noise.  As an application of the results
of Section~\ref{sec:ex1}, Section~\ref{sec:Taylor} precisely characterizes the information
storage capacity of a memory built from unreliable components.  
Section~\ref{sec:conc} provides some conclusions.

\section{Background}
\label{sec:background}

\subsection{Practical Motivations}
Although always present \cite{Hamming1950,LarssonS1994}, recent technological trends in digital circuit design bring 
practical motivations to the fore \cite{HoMH2001,ZhaoBD2007,RejimonLB2009}.  
The 2008 update of the International Technology Roadmap for
Semiconductors (ITRS)\footnote{The overall objective of the ITRS is to present the
consensus of the semiconductor industry on the best current estimate of research and development 
needs for the next fifteen years.} points out that for complementary metal-oxide-silicon (CMOS)
technology, increasing power densities, decreasing supply voltages, and decreasing sizes have
increased sensitivity to cosmic radiation, electromagnetic interference, and thermal fluctuations.
The ITRS further says that an ongoing shift in the manufacturing paradigm
will dramatically reduce costs but will lead 
to more transient failures of signals, logic values, devices, and interconnects.
Device technologies beyond CMOS, such as single-electron tunnelling
technology \cite{Likharev1999}, carbon-based nanoelectronics \cite{BachtoldHND2001}, and
chemically assembled electronic nanocomputers \cite{CollierWBRSKWH1999}, are also projected 
to enter production, but they all display erratic, random device behavior 
\cite{HanJ2003,AnghelN2007}.

Analog computations are always subject to noise \cite{Bush1931,Sarpeshkar1998}.  
Similar issues arise when performing real-valued computations on digital computers since 
quantization, whether fixed-point or floating-point, is often well-modeled 
as bounded, additive stochastic noise \cite{WidrowK2008}.

\subsection{Coding and Computing}

Information and communication theory have provided limits for processing unreliable signals 
with reliable circuits \cite{Shannon1948,Gallager1963,RichardsonU2001}, whereas
fault-tolerant computing theory has provided limits for processing
reliable signals (inputs) with unreliable circuits 
\cite{VonNeumann1956,WinogradC1963,Pierce1965,Hadjicostis2002,GacsG1994,Elias1958}.
This work brings the two together.

A brief overview of terms and concepts from fault-tolerant computing, based 
on \cite{Johnson1989,Pradhan1996}, is now provided.  A \emph{fault} is a
physical defect, imperfection, or flaw that occurs within some hardware or software component.
An \emph{error} is the informational manifestation of a fault.  A \emph{permanent}
fault exists indefinitely until corrective action is taken, whereas a \emph{transient}
fault appears and disappears in a short period of time.  Noisy circuits in which 
the interconnection pattern of components are trees are called \emph{formulas} 
\cite{Pippenger1988,EvansS1999}.

In an \emph{error model},
the effects of faults are given directly in the informational universe.  For example, the basic 
von Neumann model of noisy circuits \cite{VonNeumann1956} models transient faults
in logic gates and wires as message and node computation noise that is both spatially 
and temporally independent; this has more recently also been called the Hegde-Shanbhag model \cite{BeniniM2006}, after \cite{HegdeS2000}.  This error model is used here.  
Error models of permanent faults \cite{CrickP2003,YeungC2008} or of miswired circuit interconnection
\cite{WinogradC1963,Wei2003} have been considered elsewhere.  Such permanent errors in decoding circuits
may be interpreted as either changing the factor graph used for decoding 
or as introducing new potentials into the factor graph; the code used by 
the encoder and the code used by the decoder are different.

There are several design philosophies to combat faults.  \emph{Fault avoidance} seeks 
to make physical components more reliable.  \emph{Fault masking} seeks to
prevent faults from introducing errors.  \emph{Fault tolerance} is the ability of
a system to continue performing its function in the presence of faults.  
This paper is primarily concerned with fault tolerance,
but Section~\ref{sec:Taylor} considers fault masking. 

\subsection{Related Work}

Empirical characterizations of message-passing decoders have demonstrated 
that probability of error performance does not change much when messages are quantized at high resolution \cite{Gallager1963}.  
Even algorithms that are coarsely quantized versions of optimal belief propagation 
show little degradation in performance \cite{PingL2000,RichardsonU2001,SinghalCM2005,MiladinovicF2005,ChenDEFH2005,ZhaoZB2005,YuLST2007}.
It should be emphasized, however, that fault-free, quantized decoders 
differ significantly from decoders that make random errors.\footnote{Randomized 
algorithms \cite{MotwaniR1995} and stochastic 
computation \cite{Ribeiro1967} (used for decoding in \cite{TehraniGM2006})
make use of randomness to increase functionality, but the randomness is 
deployed in a controlled manner.}
The difference is
similar to that between control systems with finite-capacity noiseless channels 
and control systems with noisy channels of equal capacity \cite{SahaiM2006}.
Seemingly the only previous work on message-passing 
algorithms with random errors is \cite{IhlerFW2005},
which deals with problems in distributed inference.\footnote{If the graphical model 
of the code and the graph of noisy communication links
in a distributed system coincide, then the distributed
inference problem and the message-passing decoding problem can be made to coincide.} 

The information theoretic problem of mismatch capacity \cite{GantiLT2000} and its analog
for iterative decoding \cite{SaeediB2007} deal with scenarios where an incorrect
decoding metric is used.  This may arise, e.g., due to incorrect estimation of
the channel noise power.  For message-passing decoding algorithms, mismatch leads to incorrect
parameters for local computations.  These are permanent faults
rather than the kind of transient faults considered in this paper.

Noisy LDPC decoders were previously analyzed in the context of
designing reliable memories from unreliable components \cite{Taylor1968,Kuznetsov1973}
(revisited in Section~\ref{sec:Taylor}), using Gallager's original methods \cite{Gallager1963}.
Several LPDC code analysis tools have since been developed, 
including simulation \cite{MackayN1997},
expander graph arguments \cite{SipserS1996,BurshteinM2001},
EXIT charts \cite{tenBrink1999,ArdakaniK2004}, and
density evolution \cite{LubyMSS2001,RichardsonU2001,RichardsonU2004}.
This work generalizes asymptotic characterizations 
developed by Richardson and Urbanke for noiseless 
decoders \cite{RichardsonU2001}, showing that density evolution is applicable to faulty decoders. 
Expander graph arguments have also been extended to the case of 
noisy decoding in a paper \cite{ChilappagariV2007} that appeared 
concurrently with the first presentation of this work \cite{Varshney2007}.  
Note that previous works have not even considered the possibility 
that Shannon reliability is achievable with noisy decoding.

\section{Codes, Decoders, and Performance}
\label{sec:notation}
\begin{figure}
  \centering
  \includegraphics[width=3.5in]{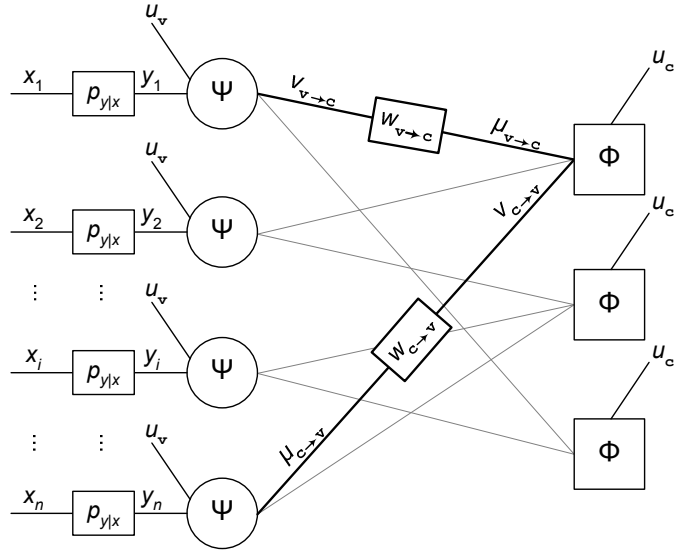}
  \caption{Schematic diagram of a factor graph-based implementation of a noisy decoder circuit.  Only one variable-to-check message and
	one check-to-variable message are highlighted.  Other wires, shown in gray, will also carry noisy messages.}
  \label{fig:factorgraph}
\end{figure}

This section establishes the basic notation of LDPC channel codes and message-passing 
decoders for communication systems depicted in Fig.~\ref{fig:fig1_rev}.  It primarily follows
established notation in the field \cite{RichardsonU2001,RichardsonU2008}, and will therefore
be brief.  Many of the notational conventions are depicted schematically in Fig.~\ref{fig:factorgraph}
using a factor graph-based decoder implementation.

Consider the standard ensemble of $(\tl,\tr)$-regular LDPC codes
of length $n$, $\mathcal{C}^n(\tl,\tr)$, defined by a uniform measure 
on the set of labeled bipartite factor graphs with variable node degree 
$\tl$ and check node degree $\tr$.\footnote{A factor graph determines
an ``ordered code,'' but the opposite is not true \cite{HalfordC2008}.  Moreover, since codes
are unordered objects, several ``ordered codes'' are in fact the same code.}  There are $n$ variable nodes
corresponding to the codeword letters and $n\tl / \tr$ check nodes
corresponding to the parity check constraints.  The design rate 
of the code is $1 - \tl/\tr$, though the actual rate might be 
higher since not all checks may be independent; the true rate
converges to the design rate for large $n$ \cite[Lemma 3.22]{RichardsonU2008}.  
One may also consider irregular codes, $\mathcal{C}^n(\lambda,\rho)$
characterized by the degree distribution pair $(\lambda,\rho)$.
Generating functions of the variable node and check node degree distributions, $\lambda(\zeta)$
and $\rho(\zeta)$, are functions of the form $\lambda(\zeta) = \sum_{i=2}^{\infty}\lambda_i \zeta^{i-1}$
and $\rho(\zeta) = \sum_{i=2}^{\infty}\rho_i \zeta^{i-1}$, where $\lambda_i$ and $\rho_i$
specify the fraction of edges that connect to nodes with degree $i$.  The design rate is
$1 - \int_0^1 \rho(\zeta)d\zeta / \int_0^1 \lambda(\zeta)d\zeta$.

In the communication system of Fig.~\ref{fig:fig1_rev}, a codeword is selected by
the transmitter and is sent through the noisy channel.
Channel input and output letters are denoted $X \in \mathcal{X}$ and 
$Y \in \mathcal{Y}$.  Since binary linear codes are used,
$\mathcal{X}$ can be taken as $\{\pm 1\}$.  The receiver contains a noisy 
message-passing decoder, which is used to process the channel output codeword
to produce an estimate of $X$ that is denoted $\hat{X}$.
The goal of the receiver is to recover the channel input codeword with low
probability of error.  Throughout this work, probability of bit error 
$P_e$ is used as the performance criterion;\footnote{An alternative would be to 
consider block error probability, however an exact evaluation of
this quantity is difficult due to the dependence between different symbols 
of a codeword, even if the bit error probability is the same
for all symbols in the codeword \cite{LentmaierTZC2005}.}
\[
P_e = \Pr[X \neq \hat{X}] \mbox{.}
\]

The message-passing decoder works in iterative stages and the iteration
time is indexed by $\ell = 0,1,\ldots$.
Within the decoder, at time $\ell = 0$,
each variable node has a realization of $Y$, $y_i$. 
A message-passing decoder exchanges messages
between nodes along wires.  First each variable node
sends a message to a neighboring check node over a noisy messaging
wire.  Generically, sent messages are denoted as $\nu_{\vtoc}$, message 
wire noise realizations as $w_{\vtoc}$, and received messages as $\mu_{\vtoc}$: assume without loss of generality
that $\nu_{\vtoc}$, $w_{\vtoc}$, and $\mu_{\vtoc}$ are drawn from a common messaging alphabet $\mathcal{M}$.  

Each check node processes
received messages and sends back a message to each neighboring
variable node over a noisy message wire.  The noisiness of the check node processing 
is generically denoted by an input random variable $U_{\tt{c}} \in \mathcal{U}$.  
The check node computation is denoted $\Phi^{(\ell)}:\mathcal{M}^{\tr-1}\times\mathcal{U} \mapsto \mathcal{M}$. 
The notations $\nu_{\ctov}$, $\mu_{\ctov}$, and $w_{\ctov}$ are used for signaling from check node to variable node;
again without loss of generality assume that $\nu_{\ctov},w_{\ctov},\mu_{\ctov} \in \mathcal{M}$.

Each variable node now processes its $y_i$ and the messages it receives
to produce new messages.  The new messages are produced
through possibly noisy processing, where the noise input is  
generically denoted $U_{\tt{v}}\in\mathcal{U}$.  
The variable node computation is denoted $\Psi^{(\ell)}:\mathcal{Y}\times\mathcal{M}^{\tl-1}\times\mathcal{U} \mapsto \mathcal{M}$.
Local computations and message-passing continue iteratively.  

Message passing induces \emph{decoding neighborhoods}, 
which involve nodes/wires that have communicated with one another. 
For a given node $\dot{n}$, its \emph{neighborhood of depth $d$}
is the induced subgraph consisting of all nodes reached and edges
traversed by paths of length at most $d$ starting from $\dot{n}$ (including
$\dot{n}$) and is denoted $\mathcal{N}_{\dot{n}}^d$.  The \emph{directed neighborhood of depth $d$} of a wire $\vtoc$, denoted
by $\mathcal{N}_{\vtoc}^d$, is defined as the induced subgraph containing all
wires and nodes on paths starting from the same place as $\vtoc$ but
different from $\vtoc$.  Equivalently for a wire $\ctov$, $\mathcal{N}_{\ctov}^d$ is 
the induced subgraph containing all wires and nodes on paths starting from 
the same place as $\ctov$ but different from $\ctov$.
If the induced subgraph (corresponding to a neighborhood) is a tree
then the neighborhood is \emph{tree-like}, otherwise it is not tree-like. 
The neighborhood is tree-like if and only if all involved nodes 
are distinct.

Note that only extrinsic information is used in node computations.  Also note
that in the sequel, all decoder noises ($U_{\tt{c}}$, $U_{\tt{v}}$, $W_{\vtoc}$, and $W_{\ctov}$) will
be assumed to be independent of each other, as in the von Neumann error model 
of faulty computing.  

A communication system is judged by information rate, error probability, and
blocklength.  For fixed channels, information theory specifies the limits of these 
three parameters when optimizing over the unconstrained choice of codes and decoders;
Shannon reliability is achievable for rates below capacity in the limit of increasing blocklength.  
When decoders are restricted to be noisy, tighter information theoretic limits
are not known.  Therefore comparing performance of systems with noisy decoders to 
systems using identical codes but noiseless decoders is more appropriate than comparing
to Shannon limits.  

Coding theory follows from information theory by restricting decoding complexity;
analysis of noisy decoders follows from coding theory by restricting decoding reliability.

\section{Density Evolution Concentration Results}
\label{sec:tools}
Considering the great successes achieved by analyzing the 
noiseless decoder performance of ensembles of codes \cite{LubyMSS2001,RichardsonU2001,RichardsonU2008}
rather than of particular codes \cite{Gallager1963}, the same approach is pursued 
for noisy decoders.  The first mathematical contribution of this work is
to extend the method of analysis promulgated in \cite{RichardsonU2001}
to the case of decoders with random noise.  

Several facts that simplify performance analysis are proven.  First, under certain symmetry conditions with wide 
applicability, the probability
of error does not depend on which codeword is transmitted.  Second,
the individual performances of codes in an ensemble are, with high probability, 
the same as the average performance of the ensemble.  Finally, this
average behavior converges to the behavior of a code defined on a 
cycle-free graph.  Performance analysis then reduces to determining 
average performance on an infinite tree: a noisy formula is analyzed
in place of general noisy circuits.

For brevity, only regular LDPC codes are considered in this section, 
however the results can be generalized to irregular LDPC codes.  In particular, 
replacing node degrees by maximum node degrees, the proofs stand \emph{mutatis mutandis}.  
Similarly, only binary LDPC codes are considered; generalizations to non-binary
alphabets also follow, as in \cite{RathiU2005}.

\subsection{Restriction to All-One Codeword}
If certain symmetry conditions are satisfied by the system, then 
the probability of error is conditionally independent of the
codeword that is transmitted.  It is assumed throughout this section
that messages in the decoder are in \emph{belief format}.
\begin{defin}
A message in an iterative message-passing decoder for a binary code
is said to be in \emph{belief format} if the sign
of the message indicates the bit estimate and the magnitude of the message
is an increasing function of the confidence level.  In particular, a positive-valued
message indicates belief that a bit is $+1$ whereas a negative-valued message indicates
belief that a bit is $-1$.  A message of magnitude $0$ indicates complete uncertainty whereas
a message of infinite magnitude indicates complete confidence in a bit value.
\end{defin}
Note, however, that it is not obvious that
this is the best format for noisy message-passing 
\cite[Appendix B.1]{RichardsonU2008}.  The symmetry conditions
can be restated for messages in other formats.

The several symmetry conditions are:
\begin{defin}[Channel Symmetry]
\label{def:def1}
A memoryless channel is binary-input output-symmetric if it satisfies
\[
p(Y_t = y|X_t = 1) = p(Y_t = -y|X_t = -1)
\]
for all channel usage times $t = 1,\ldots,n$.
\end{defin}
\begin{defin}[Check Node Symmetry]
\label{def:def2}
A check node message map is symmetric if it satisfies
\[
\Phi^{(\ell)}(b_1\mu_1,\ldots,b_{\tr - 1}\mu_{\tr - 1},b_{\tr}u) = \Phi^{(\ell)}(\mu_1,\ldots,\mu_{\tr - 1},u)\left(\prod\limits_{i=1}^{\tr}b_i\right)
\]
for any $\pm 1$ sequence $(b_1,\ldots,b_{\tr})$.  That is to say, 
the signs of the messages and the noise factor out of the map.  
\end{defin}
\begin{defin}[Variable Node Symmetry]
\label{def:def3}
A variable node message map is symmetric if it satisfies
\[
\Psi^{(0)}(-\mu_0,-u) = -\Psi^{(0)}(\mu_0,u)
\]
and
\[
\Psi^{(\ell)}(-\mu_0,-\mu_1,\ldots,-\mu_{\tl -1},-u) = -\Psi^{(\ell)}(\mu_0,\mu_1,\ldots,\mu_{\tl -1},u) \mbox{,} 
\]
for $\ell \ge 1$.  That is to say, the initial message from the 
variable node only depends on the received value and internal
noise and there is sign inversion invariance for all messages.  
\end{defin}
\begin{defin}[Message Wire Symmetry]
\label{def:def4}
Consider any message wire to be a
mapping $\Xi: \mathcal{M}\times\mathcal{M} \to \mathcal{M}$.
Then a message wire is symmetric if
\[
\mu = \Xi(\nu,w) = -\Xi(-\nu,-w)\mbox{,}
\]
where $\mu$ is any message received at a node when the message
sent from the opposite node is $\nu$ and $w$ is message wire noise
with distribution symmetric about $0$.
\end{defin}
An example where the message wire symmetry condition holds is if the message wire noise $w$ is additive
and symmetric about $0$.  Then $\mu = \nu + w = -(-\nu - w)$ and $w$ is symmetric about $0$.

\begin{thm}[Conditional Independence of Error]
\label{thm:symmetry} 
For a given binary linear code and a given noisy message-passing algorithm, 
let $P_e^{(\ell)}(\bf{x})$ 
denote the conditional probability of error after the $\ell$th decoding
iteration, assuming that codeword $\bf{x}$ was sent. If the channel and
the decoder satisfy the symmetry conditions given in Definitions~\ref{def:def1}--\ref{def:def4}, then 
$P_e^{(\ell)}(\bf{x})$ does not depend on $\bf{x}$.
\end{thm}
\begin{IEEEproof}
Modification of \cite[Lemma 1]{RichardsonU2001} or \cite[Lemma 4.92]{RichardsonU2008}.
Appendix~\ref{app:symmetry} gives details.
\end{IEEEproof}

Suppose a system meets these symmetry conditions.  
Since probability of error is independent of the transmitted codeword
and since all LDPC codes have the all-one codeword in the codebook, one may assume without 
loss of generality that this codeword is sent.   
Doing so removes the randomness associated with transmitted codeword selection.

\subsection{Concentration around Ensemble Average}
The next simplification follows by seeing that the average performance of
the ensemble of codes rather than the performance of a particular code 
may be studied, since all codes in the ensemble perform similarly. 
The performances of almost all LDPC codes closely match
the average performance of the ensemble from which they are drawn.  The average 
is over the instance of the code, the realization of the channel noise, and 
the realizations of the two forms of decoder noise.
To simplify things, assume that the number of decoder iterations is fixed 
at some finite $\ell$.  Let $Z$ be the
number of incorrect values held among all $\tl n$ variable node-incident edges
at the end of the $\ell$th iteration (for a particular code, channel noise realization, and
decoder noise realization) and let $\E{Z}$ be the expected
value of $Z$.  By constructing a martingale 
through sequentially revealing all of the random elements and then using the Hoeffding-Azuma inequality,
it can be shown that:
\begin{thm}[Concentration Around Expected Value]
\label{thm:conc}
There exists a positive constant $\beta = \beta(\tl,\tr,\ell)$ such that
for any $\epsilon > 0$, 
\[
\Pr\left[|Z-\E{Z}| > n \tl \epsilon / 2\right] \le 2e^{-\beta \epsilon^2 n} \mbox{.}
\]
\end{thm}
\begin{IEEEproof}
Follows the basic ideas of the proofs of \cite[Theorem 2]{RichardsonU2001} or \cite[Theorem 4.94]{RichardsonU2008}.
Appendix~\ref{app:concentrate} gives details.
\end{IEEEproof}

A primary communication system performance criterion is probability of error $P_e$;
if the number of incorrect values $Z$ concentrates, then so does $P_e$.   

\subsection{Convergence to the Cycle-Free Case}
The previous theorem showed that the noisy decoding algorithm behaves essentially
deterministically for large $n$.  As now shown, this ensemble 
average performance converges to the performance of an associated tree ensemble,
which will allow the assumption of independent messages.

For a given edge whose directed neighborhood
of depth $2\ell$ is tree-like, let $p$ be the expected number of incorrect
messages received
along this edge (after message noise) at the $\ell$th iteration, 
averaged over all graphs, inputs and decoder noise realizations of both types.  

\begin{thm}[Convergence to Cycle-Free Case]
\label{thm:conv}
There exists a positive constant $\gamma = \gamma(\tl,\tr,\ell)$ 
such that for any $\epsilon > 0$ and $n > 2\gamma/ \epsilon$,
\[
|\E{Z} - n\tl p| < n\tl \epsilon /2 \mbox{.}
\]
\end{thm}

The proof is identical to the proof of \cite[Theorem 2]{RichardsonU2001}.
The basic idea is that the computation tree created by unwrapping the code graph to 
a particular depth \cite{Weiss2000} almost surely has no repeated nodes.

The concentration and convergence results directly imply concentration
around the average performance of a tree ensemble:
\begin{thm}[Concentration Around Cycle-Free Case]
There exist positive constants $\beta = \beta(\tl,\tr,\ell)$ and 
$\gamma = \gamma(\tl,\tr,\ell)$ such that for any 
$\epsilon > 0$ and $n > 2\gamma/ \epsilon$,
\[
\Pr\left[|Z-n\tl p| > n \tl \epsilon\right] \le 2e^{-\beta \epsilon^2 n} \mbox{.}
\]
\end{thm}
\begin{IEEEproof}
Follows directly from Theorems \ref{thm:conc} and \ref{thm:conv}.
\end{IEEEproof}

\subsection{Density Evolution}
With the conditional independence and concentration results, all randomness is 
removed from explicit consideration and all messages are independent.  The
problem reduces to density evolution, the analysis of a discrete-time 
dynamical system \cite{RichardsonU2004}.  The dynamical system state variable 
of most interest is the probability of bit error, $P_e$. 

Denote the probability of bit error of a code $g \in \mathcal{C}^n$ after 
$\ell$ iterations of decoding by $P_e^{(\ell)}(g,\varepsilon,\alpha)$, where $\varepsilon$ is
a channel noise parameter (such as noise power or crossover probability)
and $\alpha$ is a decoder noise parameter (such as logic gate error probability).  
Then density evolution computes
\[
\lim_{n\to\infty} \E{P_e^{(\ell)}(g,\varepsilon,\alpha)}\mbox{,}
\]
where the expectation is over the choice of the code and the various noise
realizations.  The main interest is in the long-term behavior of the 
probability of error after performing many iterations.  The long-term behavior 
of a generic dynamical system may be a limit cycle or a chaotic attractor, 
however density evolution usually converges to a stable fixed 
point.  Monotonicity (either increasing or decreasing) with respect to 
iteration number $\ell$ need not hold, but it often does.
If there is a stable fixed point, the limiting performance corresponds to
\[
\eta^* = \lim_{\ell\to\infty}\lim_{n\to\infty} \E{P_e^{(\ell)}(g,\varepsilon,\alpha)}\mbox{.}
\]
In channel coding, certain sets of parameters 
$(g,\varepsilon,\alpha)$ lead to ``good'' performance, in the sense
of small $\eta^*$, whereas other sets of parameters lead
to ``bad'' performance with large $\eta^*$.  The goal
of density evolution analysis is to determine the boundary between 
these good and bad sets.  

Though it is natural to expect the performance of an algorithm 
to improve as the quality of its input improves and as more resources
are allocated to it, this may not be so.
For many decoders, however, there is a monotonicity 
property that limiting behavior $\eta^*$ improves as channel noise 
$\varepsilon$ decreases and as decoder noise $\alpha$ decreases.
Moreover, just as in other nonlinear estimation systems for
dimensionality-expanding signals \cite{Woodward1953,Slepian1962,WozencraftJ1965}, 
there is a threshold phenomenon such that the limiting probability of error
may change precipitously with the values of $\varepsilon$ and $\alpha$. 

In traditional coding theory, there is no parameter $\alpha$, and
the goal is often to determine the range of $\varepsilon$ for 
which $\eta^*$ is zero.  The boundary is often called the decoding 
threshold and may
be denoted $\varepsilon^*(\eta^* = 0)$.  A decoding threshold
for optimal codes under optimal decoding may be computed from the
rate of the code $g$ and the capacity of the channel as a function 
of $\varepsilon$, $C(\varepsilon)$.  Since
this Shannon limit threshold is for optimal codes and decoders, it is clearly an 
upper bound to $\varepsilon^*(0)$ for any given code and decoder.
If the target error probability $\eta^*$ is non-zero, then the Shannon
limit threshold is derived from the so-called $\eta^*$-capacity, 
$\tfrac{C(\varepsilon)}{1-h_2(\eta^*)}$, rather than $C(\varepsilon)$.\footnote{The 
function $h_2(\cdot)$ is the binary entropy function.  The 
$\eta^*$-capacity expression is obtained by adjusting capacity by the rate-distortion 
function of an equiprobable binary source under frequency 
of error constraint $\eta^*$, $R(\eta^*) = 1 - h_2(\eta^*)$ \cite{Erokhin1958}.}

In the case of faulty decoders, the Shannon limits also provide upper bounds 
on the $\varepsilon$-boundary for the set of $(\varepsilon,\alpha)$ that 
achieve good performance.  One might hope for a Shannon theoretic characterization
of the entire $(\varepsilon,\alpha)$-boundary, but as noted previously, such results
are not extant.  Alternately, in the next sections, sets of $(\varepsilon,\alpha)$ 
that can achieve $\eta^*$-reliability for particular LDPC codes $g\in \mathcal{C}^n$
are characterized using the density evolution method developed in this section. 

\section{Example: Noisy Gallager A Decoder}
\label{sec:ex1}
Section \ref{sec:tools} showed that density 
evolution equations determine the performance of almost all
codes in the large blocklength regime.  Here the density evolution
equation for a simple noisy message-passing decoder, a noisy
version of Gallager's decoding algorithm A 
\cite{Gallager1963,BazziRU2004}, is derived.  
The algorithm has message alphabet $\mathcal{M} = \{\pm 1\}$, with messages in belief format
simply indicating the estimated sign of a bit.  Although this simple decoding algorithm cannot match the performance of 
belief propagation due to its restricted messaging alphabet $\mathcal{M}$, 
it is of interest since it is of extremely low complexity and can be analyzed analytically \cite{BazziRU2004}. 

Consider decoding the LDPC-coded output of a binary symmetric channel (BSC) 
with crossover probability $\varepsilon$.
At a check node, the outgoing message along edge $\vec{e}$ is the product of all
incoming messages excluding the one incoming on $\vec{e}$, i.e.\ the check node map $\Phi$ is the XOR operation.
At a variable node, the outgoing message is the original received code symbol unless all incoming messages
give the opposite conclusion.  That is,
\[
\Psi = \begin{cases} -y\mbox{,} & \mbox{if } \mu_1 = \cdots = \mu_{\tl-1} = -y \mbox{,}\\ y\mbox{,} & \mbox{otherwise.} \end{cases}
\]

\begin{figure}
  \centering
	\begin{tabular}{c}
  \includegraphics[width=2.5in]{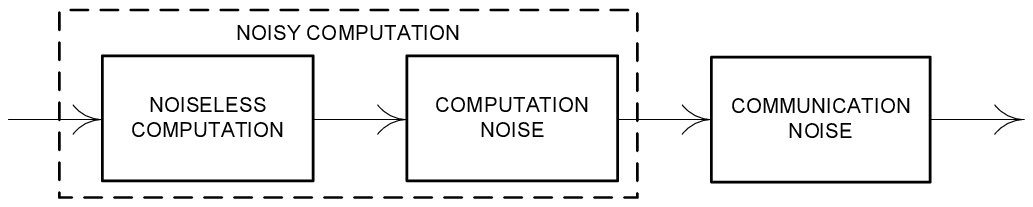} \\
  \Large{$\approx$} \\ 
  \includegraphics[width=2.5in]{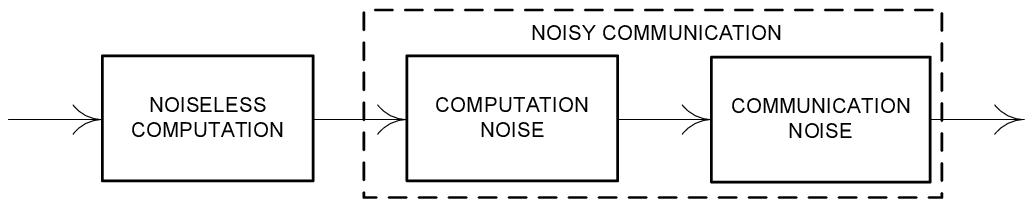}
	\end{tabular}
  \caption{Local computation noise may be incorporated into message-passing noise without essential loss of generality.}
  \label{fig:combinenoise}
\end{figure}

There is no essential loss
of generality by combining computation noise and message-passing
noise into a single form of noise, as demonstrated schematically in Fig.~\ref{fig:combinenoise} and proven in
\cite[Lemma 3.1]{DobrushinO1977}.  This noise combining is performed in the sequel to reduce the number
of decoder noise parameters and allow a clean examination
of the central phenomenon.  Thus, each message in the Gallager algorithm A
is passed over an independent and identical BSC wire with crossover probability $\alpha$.  

The density evolution equation leads to an analytic 
characterization of the set of $(\varepsilon,\alpha)$ pairs,
which parameterize the noisiness of the communication system.

\subsection{Density Evolution Equation}
The density evolution equation
is developed for general irregular LDPC ensembles.
The state variable of density evolution, $s_{\ell}$, is 
taken to be the expected probability of bit error 
at the variable nodes in the large blocklength limit, denoted here as $P_{e}^{(\ell)}(\varepsilon,\alpha)$.

The original received message is in error with probability $\varepsilon$, 
thus 
\[
P_{e}^{(0)}(\varepsilon,\alpha) = s_0 = \varepsilon \mbox{.}
\]

The initial variable-to-check message is in error with probability 
$(1-\varepsilon)\alpha + \varepsilon(1-\alpha)$,
since it is passed through a BSC($\alpha$).  For further iterations, $\ell$, the probability of error, 
$P_{e}^{(\ell)}(\varepsilon,\alpha)$, is found by induction.  Assume
$P_{e}^{(i)}(\varepsilon,\alpha) = s_{i}$  for $0\le i\le\ell$.  Now consider
the error probability of a check-to-variable message in the $(\ell + 1)$th iteration.
A check-to-variable message emitted by a check node of degree $\tr$ along a
particular edge is the product of all the $(\tr-1)$ incoming messages along
all other edges.  By assumption, each such message is in error with probability
$s_{\ell}$ and all messages are independent.  These messages are passed through BSC($\alpha$)
before being received, so the probability of being received in error is
\[
s_{\ell}(1-\alpha) + (1-s_{\ell})\alpha = \alpha + s_{\ell} - 2\alpha s_{\ell} \mbox{.}
\]

Due to the XOR operation, the outgoing message will be in error if an odd number of these received messages are in error.
The probability of this event, averaged over the degree distribution, yields the probability
\[
\frac{1 - \rho\left[1 - 2( \alpha + s_{\ell} - 2\alpha s_{\ell})\right]}{2} \mbox{.}
\]

Now consider $P_{e}^{(\ell+1)}(\varepsilon,\alpha)$, the error probability at the variable
node in the $(\ell + 1)$th iteration.  Consider an edge which is connected to a variable node of degree $\tl$.
The outgoing variable-to-check message along this edge is in error in the $(\ell + 1)$th iteration
if the original received value is in error and not all incoming messages are received correctly or if the originally 
received value is correct but all incoming messages are in error.  The first event has probability
\[
\varepsilon\left(1-\left[1-(1-\alpha)\left(\frac{1 - \rho\left[1 - 2( \alpha + s_{\ell} - 2\alpha s_{\ell})\right]}{2}\right) - \alpha\left(\frac{1 + \rho\left[1 - 2( \alpha + s_{\ell} - 2\alpha s_{\ell})\right]}{2}\right)\right]^{\tl-1}\right) \mbox{.}
\]
The second event has probability
\[
(1-\varepsilon)\left(\left[(1-\alpha)\left(\frac{1 - \rho\left[1 - 2( \alpha + s_{\ell} - 2\alpha s_{\ell})\right]}{2}\right) + \alpha\left(\frac{1 + \rho\left[1 - 2( \alpha + s_{\ell} - 2\alpha s_{\ell})\right]}{2}\right)\right]^{\tl-1}\right) \mbox{.}
\]

Averaging over the degree distribution and adding the two terms together 
yields the density evolution equation in recursive form:
\begin{equation}
s_{\ell + 1} = \varepsilon - \varepsilon q_{\alpha}^{+}(s_{\ell}) + (1-\varepsilon) q_{\alpha}^{-}(s_{\ell}) \mbox{.}
\label{eq:DE}
\end{equation}
The expressions
\[
q_{\alpha}^{+}(\check{s}) = \lambda\left[\frac{1 + \rho(\omega_{\alpha}(\check{s})) - 2\alpha\rho(\omega_{\alpha}(\check{s}))}{2}\right] \mbox{,}
\]
\[
q_{\alpha}^{-}(\check{s}) = \lambda\left[\frac{1 - \rho(\omega_{\alpha}(\check{s})) + 2\alpha\rho(\omega_{\alpha}(\check{s}))}{2}\right] \mbox{,}
\]
and $\omega_{\alpha}(\check{s}) = (2\alpha - 1)(2\check{s} - 1)$ are used to define the density evolution recursion.

\subsection{Performance Evaluation}
With the density evolution equation established, the performance
of the coding-decoding system with particular values of quality parameters 
$\varepsilon$ and $\alpha$ may be determined.  
Taking the bit error probability as the state variable, stable 
fixed points of the deterministic, discrete-time, dynamical
system are to be found.  Usually
one would want the probability of error to converge to zero, but since this might not be possible, 
a weaker performance criterion may be needed.  To start, consider partially noiseless cases.
 
\subsubsection{Noisy Channel, Noiseless Decoder}
For the noiseless decoder case, i.e.\ $\alpha = 0$, it has been known that there 
are thresholds on $\varepsilon$, below which the probability of error goes to zero 
as $\ell$ increases, and above which the probability of error 
goes to some large value.  These can be found analytically for the Gallager 
A algorithm \cite{BazziRU2004}.  

\subsubsection{Noiseless Channel, Noisy Decoder}
For the noisy Gallager A system under consideration, the probability of error 
does not go to zero as $\ell$ goes to infinity for 
any $\alpha > 0$.  This can be seen by considering the case of 
the perfect original channel, $\varepsilon = 0$, and any $\alpha > 0$.
The density evolution equation reduces to 
\begin{equation}
s_{\ell + 1} = q_{\alpha}^{-}(s_{\ell}) \mbox{,}
\label{eq:DEorignoiseless}
\end{equation}
with $s_0 = 0$.  The recursion does not have a fixed point at zero, and since error probability is 
bounded below by zero, it must increase.  The derivative is
\[
\frac{\partial}{\partial s} q_{\alpha}^{-}(s) = \lambda^{\prime}\left[\frac{1-\rho(\omega_{\alpha}(s))+2\alpha\rho(\omega_{\alpha}(s))}{2}\right]\rho^{\prime}(\omega_{\alpha}(s))(2\alpha-1)^2 \mbox{,}
\]
which is greater than zero for $0\le s\le \tfrac{1}{2}$ and $0\le\alpha\le \tfrac{1}{2}$; 
thus the error evolution forms a monotonically increasing sequence.  Since the sequence 
is monotone increasing starting from zero, and there is no fixed point at zero, it follows 
that this converges to the smallest real solution of $s = q_{\alpha}^{-}(s)$ since the fixed point 
cannot be jumped due to monotonicity.  

\subsubsection{Noisy Channel, Noisy Decoder}
The same phenomenon must also happen if the starting $s_0$ is positive, however the 
value to which the density evolution converges is a non-zero fixed point solution of the 
original equation (\ref{eq:DE}), not of (\ref{eq:DEorignoiseless}), and is
a function of both $\alpha$ and $\varepsilon$.  Intuitively, for somewhat large initial values of 
$\varepsilon$, the noisy decoder decreases the probability of error in the first few iterations, just 
like the noiseless one, but when the error probability becomes close to the internal decoder 
error, the probability of error settles at that level.  This is summarized in the following proposition.
\begin{prop}
\label{prop:nonzero}
Final error probability $\eta^* > 0$ for any LDPC ensemble decoded using the noisy Gallager A system defined in Section~\ref{sec:ex1}, 
for every decoder noise level $\alpha > 0$ and every channel noise level $\varepsilon$. $\Box$
\end{prop}

The fact that probability of error cannot asymptotically be driven to zero with the
noisy Gallager decoder is expected yet is seemingly displeasing.  In a practical 
scenario, however, the ability to drive $P_e$ to a very small number
is also desirable.  As such, a performance objective of achieving $P_e$
less than $\eta$ is defined and the worst channel (ordered by $\varepsilon$) for
which a decoder with noise level $\alpha$ can achieve that objective is determined.  
The channel parameter 
\[
\varepsilon^*(\eta,\alpha) = \sup\{\varepsilon \in [0,\tfrac{1}{2}] \mid \lim_{\ell \to \infty} P_e^{(\ell)}(g,\varepsilon,\alpha) < \eta\}
\]
is called the threshold.  For a large interval of $\eta$ values, there is a single 
threshold value below which $\eta$-reliable communication is possible and above which it is not.
Alternatively, one can determine the probability of error to which a system with particular
$\alpha$ and $\varepsilon$ can be driven, $\eta^*(\alpha,\varepsilon) = \lim_{\ell \to \infty} P_e^{(\ell)}$, 
and see whether this value is small.

In order to find the threshold in the case of $\alpha > 0$ and $\varepsilon > 0$,
the real fixed point solutions of density evolution recursion \eqref{eq:DE} need to be found.
The real solutions of the polynomial equation in $s$,
\[
\varepsilon - \varepsilon q_{\alpha}^{+}(s) + (1-\varepsilon) q_{\alpha}^{-}(s) - s = 0
\]
are denoted  
$0 < r_1(\alpha,\varepsilon) \le r_2(\alpha,\varepsilon) \le r_3(\alpha,\varepsilon) \le \cdots$.\footnote{The 
number of real solutions can be determined through Descartes' rule of signs or a
similar tool \cite{Yang1999}. \label{fn:descartes}}
The final probability of error $\eta^*$ is determined by the $r_i$,
since these are fixed points of the recursion \eqref{eq:DE}.

The real solutions of the polynomial equation in $s$,
\begin{equation}
\label{eq:taueq}
\frac{s - q_{\alpha}^{-}(s)}{1 - q_{\alpha}^{+}(s) - q_{\alpha}^{-}(s)} - s = 0 \mbox{,}
\end{equation}
are denoted $0 < \tau_1(\alpha) \le \tau_2(\alpha) \le \cdots$.$^{\textrm{\ref{fn:descartes}}}$
The threshold $\varepsilon^*$ as well as the region in the $\alpha-\varepsilon$ plane where the decoder 
improves performance over no decoding are determined by the $\tau_i$, since \eqref{eq:taueq} is obtained
by solving recursion \eqref{eq:DE} for $\varepsilon$ and setting equal to zero.    For particular ensembles of LDPC codes, these
values can be computed analytically.  For these particular ensembles, it can be determined
whether the fixed points are stable or unstable.  Moreover, various monotonicity
results can be established to show that fixed points cannot be jumped.

\begin{figure}
  \centering
  \includegraphics[width=2.8in]{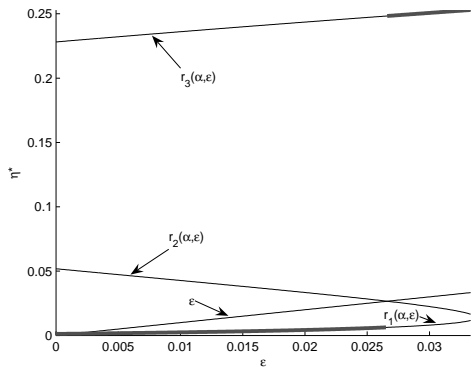}
  \caption{Thick line shows final error probability, $\eta^*$, after decoding a $\mathcal{C}^{\infty}(3,6)$ 
	code with the noisy Gallager A algorithm, $\alpha = 0.005$.  This is determined by the fixed points of 
	density evolution, $r_i(\alpha,\varepsilon)$, shown with thin lines.}
  \label{fig:fixed_alpha_plot}
\end{figure}

Analytical expressions for the $r_i(\alpha,\varepsilon)$ and $\tau_i(\alpha)$ are determined for
the (3,6) regular LDPC code by solving the appropriate polynomial equations and  
numerical evaluations of the $r_i$ expressions are shown as thin lines
in Fig.~\ref{fig:fixed_alpha_plot} as functions of $\varepsilon$ for fixed $\alpha$.
The point where $r_1(\alpha,\varepsilon) = \varepsilon$ is $\tau_1(\alpha)$ and the point 
where $r_2(\alpha,\varepsilon) = \varepsilon$ is $\tau_2(\alpha)$.  In Fig.~\ref{fig:fixed_alpha_plot}, 
these are points where the thin lines cross.

By analyzing the dynamical system equation \eqref{eq:DE} for the (3,6) code in detail, 
it can be shown that $r_1(\alpha,\varepsilon)$ and $r_3(\alpha,\varepsilon)$ are stable fixed points of density evolution. 
Contrarily, $r_2(\alpha,\varepsilon)$ is an unstable fixed point, which determines the boundary between the regions of attraction
for the two stable fixed points.  Since $r_1(\alpha,\varepsilon)$ and $r_3(\alpha,\varepsilon)$ are stable fixed points, the final
error probability $\eta^*$ will take on one of these two values, depending on the starting point of
the recursion, $\varepsilon$.  The thick line in Fig.~\ref{fig:fixed_alpha_plot} shows
the final error probability $\eta^*$ as a function of initial error probability $\varepsilon$.
One may note that $\eta^* = r_1$ is the desirable small error probability, whereas 
$\eta^* = r_3$ is the undesirable large error probability and that $\tau_2$ delimits
these two regimes.  
 
\begin{figure}
  \centering
  \includegraphics[width=2.8in]{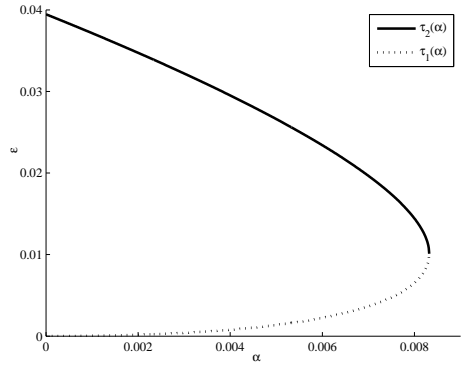}
  \caption{Decoding a $\mathcal{C}^{\infty}(3,6)$ code with the noisy Gallager A algorithm.  Region where it is beneficial to use decoder
	is below $\tau_2$ and above $\tau_1$.}
  \label{fig:region_to_use_decoder}
\end{figure}

The $\tau(\alpha)$ points determine when 
it is beneficial to use the decoder, in the sense that $\eta^* < \varepsilon$.  
By varying $\alpha$ (as if in a sequence of plots like Fig.~\ref{fig:fixed_alpha_plot}), 
an $\alpha-\varepsilon$ region where the decoder is beneficial is demarcated; this is shown in 
Fig.~\ref{fig:region_to_use_decoder}.  The function $\tau_2(\alpha)$ is the $\eta$-reliability
decoding threshold for large ranges of $\eta$.

Notice that the previously known special case, the decoding threshold of the noiseless decoder, can be recovered 
from these results.  The decoding threshold for the noiseless decoder is denoted $\varepsilon^*_{BRU}$  
and is equal to the following expression \cite{BazziRU2004}.
\[
\varepsilon^*_{BRU}  = \frac{1-\sqrt{\sigma}}{2}\mbox{,}
\]
where
\[
\sigma = -\frac{1}{4} + \frac{\sqrt{-\tfrac{5}{12}-b}}{2} + \frac{\sqrt{-\tfrac{5}{6}+\tfrac{11}{4\sqrt{-5/12-b}}}}{2}
\]
and
\[
b = \frac{8}{3}\left(\frac{2}{83 + 3\sqrt{993}}\right)^{\tfrac{1}{3}} -\frac{1}{3}\left( \frac{83 + 3\sqrt{993}}{2} \right)^{\tfrac{1}{3}} \mbox{.}
\]
This value is recovered from noisy decoder results by noting that $\eta^*(\alpha=0,\varepsilon) = 0$
for $\varepsilon \in \left[0,\varepsilon^*_{BRU}\right]$, 
which are the ordinate intercepts of the region in Fig.~\ref{fig:region_to_use_decoder}.

To provide a better sense of the performance of the noisy Gallager A algorithm, Table~\ref{tab:performance} lists
some values of $\alpha$, $\varepsilon$, and $\eta^*$ (numerical evaluations are listed and an example of an analytical expression
is given in Appendix~\ref{app:analexpr}).
As can be seen from these results, particularly from the $\tau_2$ curve 
in Fig.~\ref{fig:region_to_use_decoder}, the error probability performance of the
system degrades gracefully as noise is added to the decoder.


\begin{table}
  \caption{Performance of Noisy Gallager A algorithm for (3,6) code}
  \label{tab:performance}
  \centering
   \footnotesize
    \begin{tabular}{|r|r|r|r|}
      \hline
      \footnotesize{$\alpha$} & \footnotesize{$\varepsilon^*(0.1,\alpha)$} & \footnotesize{$\eta^*(\alpha,\varepsilon^*)$} & \footnotesize{$\eta^*(\alpha,0.01)$} \\ \hline
      \footnotesize{$0$}                & \footnotesize{$0.0394636562$} & \footnotesize{$0$} & \footnotesize{$0$}\\ \hline
      \footnotesize{$1\times 10^{-10}$} & \footnotesize{$0.0394636560$} & \footnotesize{$7.8228\times 10^{-11}$} & \footnotesize{$1.3333\times 10^{-11}$}\\ \hline
      \footnotesize{$1\times 10^{-8}$}  & \footnotesize{$0.0394636335$} & \footnotesize{$7.8228\times 10^{-9}$}  & \footnotesize{$1.3333\times 10^{-9}$}\\ \hline
      \footnotesize{$1\times 10^{-6}$}  & \footnotesize{$0.0394613836$} & \footnotesize{$7.8234\times 10^{-7}$}  & \footnotesize{$1.3338\times 10^{-7}$}\\ \hline
      \footnotesize{$1\times 10^{-4}$}  & \footnotesize{$0.0392359948$} & \footnotesize{$7.8866\times 10^{-5}$}  & \footnotesize{$1.3812\times 10^{-5}$}\\ \hline
      \footnotesize{$3\times 10^{-4}$}  & \footnotesize{$0.0387781564$} & \footnotesize{$2.4050\times 10^{-4}$}  & \footnotesize{$4.4357\times 10^{-5}$}\\ \hline
      \footnotesize{$1\times 10^{-3}$}  & \footnotesize{$0.0371477336$} & \footnotesize{$8.4989\times 10^{-4}$}  & \footnotesize{$1.8392\times 10^{-4}$}\\ \hline
      \footnotesize{$3\times 10^{-3}$}  & \footnotesize{$0.0321984070$} & \footnotesize{$3.0536\times 10^{-3}$}  & \footnotesize{$9.2572\times 10^{-4}$}\\ \hline
      \footnotesize{$5\times 10^{-3}$}  & \footnotesize{$0.0266099758$} & \footnotesize{$6.3032\times 10^{-3}$}  & \footnotesize{$2.4230\times 10^{-3}$}\\ \hline
    \end{tabular}
\end{table} 

\begin{figure}
  \centering
  \includegraphics[width=2.8in]{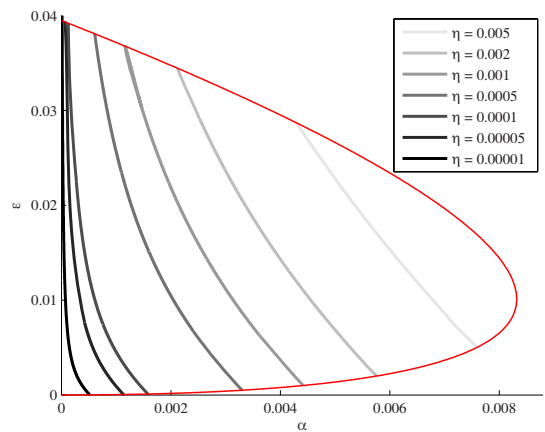}
  \caption{$\eta$-thresholds (gray lines) for decoding a $\mathcal{C}^{\infty}(3,6)$ code with the noisy Gallager A algorithm 
	within the region to use decoder (delimited with red line).}
  \label{fig:contourplot}
\end{figure}

Returning to threshold characterization, an analytical expression for the threshold 
within the region to use decoder is:
\[
\varepsilon^*(\eta,\alpha) = \frac{\eta - q_{\alpha}^{-}(\eta)}{1 - q_{\alpha}^{+}(\eta)-q_{\alpha}^{-}(\eta)} \mbox{,}
\]
which is the solution to the polynomial equation in $\check{\epsilon}$,
\[
\check{\epsilon} - \check{\epsilon} q_{\alpha}^{+}(\eta) + (1-\check{\epsilon})q_{\alpha}^{-}(\eta) - \eta = 0 \mbox{.}
\]
The threshold is drawn for several values of $\eta$ in Fig.~\ref{fig:contourplot}.
A threshold line determines the equivalence of channel noise and decoder noise
with respect to final probability of error.  If for example, the binary symmetric
channels in the system are a result of hard-detected AWGN channels, such a line may 
be used to derive the equivalent channel noise power for decoder noise power or
vice versa.  Threshold lines therefore provide guidelines for power allocation in communication systems.

\subsection{Code Optimization}
\label{sec:optim}
At this point, the bit error performance of a system has simply been measured; no
attempt has been made to optimize a code for a particular decoder and set of parameters.  
For fault-free decoding, it has been demonstrated that irregular code ensembles can perform 
much better than regular code ensembles like the (3,6) LDPC considered above 
\cite{RichardsonSU2001,BazziRU2004}.  
One might hope for similar improvements when
LDPC code design takes decoder noise into account.  The space of system 
parameters to be considered for noisy decoders is much larger than 
for noiseless decoders.

\begin{figure}
  \centering
  \includegraphics[width=3in]{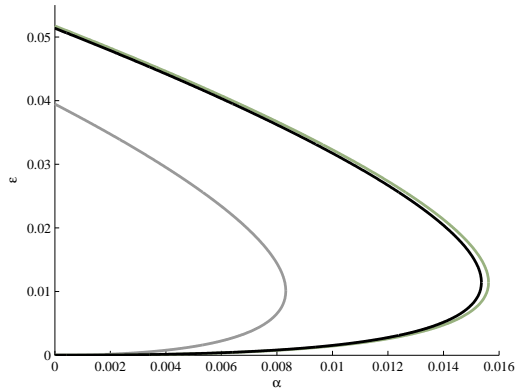}
  \caption{Region to use decoder for Bazzi et al.'s optimized rate $1/2$ LDPC code with 
	noisy Gallager A decoding (black) is contained within the region to use decoder for a rate $1/2$ LDPC code in
	Bazzi et al.'s optimal family of codes with $a = 1/10$ (green) and contains the region to use decoder 
	for the $\mathcal{C}^{\infty}(3,6)$ code (gray).
  }
  \label{fig:region}
\end{figure}

As a first step, consider the ensemble of rate $1/2$ LDPC codes that 
were optimized by Bazzi et al.\ for the fault-free Gallager A decoding algorithm
\cite{BazziRU2004}.
The left degree distribution is 
\[
\lambda(\zeta) = a\zeta^2 + (1-a)\zeta^3
\]
and the right degree distribution is
\[
\rho(\zeta) = \frac{7a}{3}\zeta^6 + \frac{3-7a}{3}\zeta^7 \mbox{,}
\]
where the optimal $a$ is specified analytically.  Numerically, $a_{\opt} = 0.1115\ldots$.  
Measuring the performance of this code with the noisy Gallager A decoder
yields the region to use decoder shown in Fig.~\ref{fig:region}; the
region to use decoder for the (3,6) code is shown for comparison.  By essentially any 
criterion of performance, this optimized code is better than the (3,6) code.

Are there other codes that can perform better on the faulty decoder than
the code optimized for the fault-free decoder?  To see whether this is possible, arbitrarily
restrict to the family of ensembles that were found to contain 
the optimal degree distribution for the fault-free decoder and take $a = 1/10$.  Also let 
$\alpha = 1/500$ be fixed.  The numerical value of the threshold 
$\varepsilon_{1/10}^*(1/10, \alpha) = 0.048239$, whereas the 
numerical value of the threshold $\varepsilon_{a_{\opt}}^*(1/10, \alpha) = 0.047857$.
In this sense, the $a = 1/10$ code is better than the $a = a_{\opt}$ code.  
In fact, as seen in Fig.~\ref{fig:region}, the region to use decoder for this $a = 1/10$ code 
contains the region to use decoder for the $a_{\opt}$ code.

On the other hand, the final error probability when operating at threshold for the $a = 1/10$ code
$\eta_{1/10}^*(\alpha,\varepsilon_{1/10}^*(1/10, \alpha)) = 0.01869$, whereas the final error probability
when operating at threshold for the $a = a_{\opt}$ code is 
$\eta_{a_{\opt}}^*(\alpha,\varepsilon_{a_{\opt}}^*(1/10, \alpha)) = 0.01766$.
So in this sense, the $a = a_{\opt}$ code is better than the $a = 1/10$ code.
The fact that highly optimized ensembles usually lead to more simultaneous 
critical points is the main complication.

If both threshold and final bit error probability are performance criteria,
there is no total order on codes and therefore there may be no notion of an optimal code.  

\section{Example: Noisy Gaussian Decoder}
\label{sec:ex2}
It is also of interest to analyze a noisy version of the belief propagation
decoder applied to the output of a continuous-alphabet channel.
Density evolution for belief propagation is difficult to 
analyze even in the noiseless decoder case, and so a Gaussian approximation method \cite{ChungRU2001}
is used.  The state variables are one-dimensional rather
than infinite-dimensional as for full analysis of belief 
propagation.  The specific node computations carried out by the decoder
are as in belief propagation \cite{RichardsonU2001}; these can be approximated by
the functions $\Phi$ and $\Psi$ defined below.  The messages and noise model are specified
in terms of the approximation.

Section~\ref{sec:ex1} had considered decoding the output of a BSC with a decoder
that was constructed with BSC components and Proposition~\ref{prop:nonzero} had 
shown that probability of bit error could never be driven to zero.  Here, the probability 
of bit error does in fact go to zero.

Consider a binary input AWGN channel with variance $\varepsilon^2$.  The output is
decoded using a noisy Gaussian decoder.  For simplicity, only regular LDPC codes 
are considered.  The messages that are passed in this decoder are real-valued,
$\mathcal{M} = \mathbb{R} \cup \{\pm \infty\}$, and are in belief format.

The variable-to-check messages in the zeroth iteration are 
the log-likelihood ratios computed from the channel output symbols, $\nu(y)$, 
\[
\nu_{\vtoc} = \nu(y) = \log \frac{p(y|x=1)}{p(y|x=-1)} \mbox{.}
\]

The check node takes the received versions of these messages,
$\mu_{\vtoc}$, as input.  The node implements a mapping $\Phi$ whose output,
$\nu_{\ctov}$, satisfies:
\[
\etanh(\nu_{\ctov}) = \prod_{i=1}^{\tr -1} \etanh(\mu_{{\vtoc}_i}) \mbox{,}
\]
where the product is taken over messages on all incoming edges except the one
on which the message will be outgoing, and  
\[
\etanh(\check{v}) = \frac{1}{\sqrt{4\pi \check{v}}}\int_{\mathbb{R}} \tanh \frac{v}{2} e^{-\frac{(v - \check{v})^2}{4v}}dv \mbox{.}
\]
The check node mapping is motivated by Gaussian likelihood computations.
For the sequel, it is useful to define a slightly different function
\[
\phi(\check{v}) = \begin{cases} 1 - \etanh(\check{v})\mbox{,}& \check{v} > 0 \\
1\mbox{,}& \check{v} = 0
\end{cases}
\]
which can be approximated as
\[
\phi(\check{v}) \approx e^{a\check{v}^c  + b} \mbox{,}
\]
with $a = -0.4527$, $b = 0.0218$, $c = 0.86$ \cite{ChungRU2001}.  

For iterations $\ell \ge 1$, the variable node takes the received versions
of the $\ctov$ messages, $\mu_{\ctov}$, as inputs.  The mapping $\Psi$ yields
output $\nu_{\vtoc}$ given by
\[
\nu_{\vtoc} = \nu(y) + \sum_{i=1}^{\tl - 1} \mu_{{\ctov}_i} \mbox{,}
\]
where the sum is taken over received messages from the neighboring check nodes
except the one to which this message is outgoing.  Again, the operation
of the variable node is motivated by Gaussian likelihood computations.

As in Section~\ref{sec:ex1}, local computation noise is combined into message-passing 
noise (Fig.~\ref{fig:combinenoise}).  To model quantization \cite{WidrowK2008}
or random phenomena, consider each message passed in the decoder to be
corrupted by signal-independent additive noise which is bounded as $-\alpha/2 \le w \le \alpha/2$.
This class of noise models includes uniform noise, and truncated
Gaussian noise, among others.  If the noise is symmetric, then Theorem~\ref{thm:symmetry} applies.  
Following the von Neumann error model, each noise realization $w$ is assumed to be 
independent.

\subsection{Density Evolution Equation}
The definition of the computation rules and the noise model may be used to derive the 
approximate density evolution equation.  The one-dimensional state variable chosen to be tracked 
is $s$, the mean belief at a variable node.  The symmetry condition relating mean belief to belief 
variance \cite{RichardsonU2001,ChungRU2001} is enforced.
Thus, if the all-one codeword was transmitted, then the value $s$ going to $+\infty$ implies that the density of 
$\nu_{\vtoc}$ tends to a ``mass point at infinity,'' which in turn implies that $P_e$ goes to $0$.

To bound decoding performance under any noise model in the class of additive
bounded noise, consider (non-stochastic) worst-case noise.  Assuming that the all-one codeword was sent, 
all messages should be as positive as possible to move towards the correct decoded codeword 
(mean beliefs of $+\infty$ indicate perfect confidence in a bit being $1$).
Consequently, the worst bounded noise that may be imposed is to subtract $\alpha/2$ 
from all messages that are passed; this requires knowledge of the transmitted codeword being all-one.
If another codeword is transmitted, then certain messages would have $\alpha/2$
added instead of subtracted.  

Such a worst-case noise model does not meet the conditions of Theorem~\ref{thm:symmetry},
but transmission of the all-one codeword is assumed nonetheless.  
If there were an adversary with knowledge of the transmitted codeword imposing worst-case noise 
on the decoder, then probability of bit error would be conditionally independent of the
transmitted codeword, as given in Appendix~\ref{app:worst}.

Note that the adversary is restricted to selecting each noise realization 
independently.  More complicated and devious error patterns in space 
or in time are not possible in the von Neumann error model.  Moreover, the
performance criterion is probability of bit error rather than probability of
block error, so complicated error patterns would provide no great benefit to the adversary.  

Since the noise is conditionally deterministic given the transmitted codeword, 
derivation of the density evolution equation is much simplified.  An 
induction argument is used, and the base case is
\[
s_0 = \tfrac{2}{\varepsilon^2} \mbox{,}
\]
where $\varepsilon^2$ is the channel noise power.  This follows from
the log-likelihood computation for an AWGN communication channel with input 
alphabet $\mathcal{X} = \{\pm 1\}$.   

The inductive assumption in the induction argument is $s_{\ell-1}$.  
This message is communicated over message-passing noise to get
\[
s_{\ell-1} - \tfrac{\alpha}{2} \mbox{.}
\]
Next the check node computation is made to yield
\[
\phi^{-1}\left(1-[1-\phi(s_{\ell-1} - \tfrac{\alpha}{2})]^{\tr-1} \right) \mbox{.}
\]
By the inductive assumption, all messages will be equivalent; that is why the product is a $(\tr-1)$-fold
product of the same quantity.  This value is communicated over message-passing noise to get
\[
\phi^{-1}\left(1-[1-\phi(s_{\ell-1} - \tfrac{\alpha}{2})]^{\tr-1} \right) - \tfrac{\alpha}{2} \mbox{.}
\]
Finally the variable-node computation yields
\[
s_0 + (\tl - 1)\left\{ \phi^{-1}\left(1-[1-\phi(s_{\ell-1} - \tfrac{\alpha}{2})]^{\tr-1} \right) - \tfrac{\alpha}{2} \right\} \mbox{.}
\]
Again, all messages will be equivalent so the sum is a $(\tl - 1)$-fold sum of
the same quantity.  Thus the density evolution equation is
\begin{equation}
\label{eq:DEgauss}
s_{\ell} = \tfrac{2}{\varepsilon^2} - \tfrac{(\tl-1)\alpha}{2} + (\tl - 1)\left\{ \phi^{-1}\left(1-[1-\phi(s_{\ell-1} - \tfrac{\alpha}{2})]^{\tr-1} \right) \right\}  \mbox{.}
\end{equation}

\subsection{Performance Evaluation}
\begin{figure}
  \centering
  \includegraphics[width=3in]{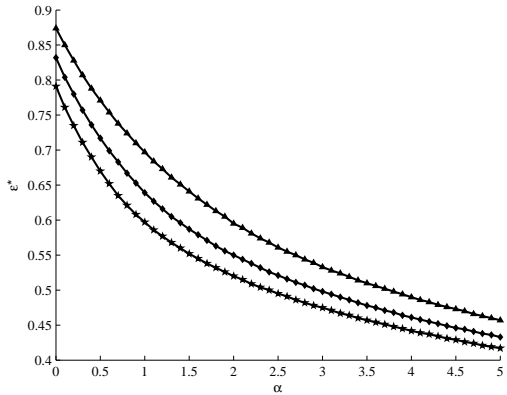}
  \caption{Thresholds for decoding the $\mathcal{C}^{\infty}(3,6)$ code (triangle), the $\mathcal{C}^{\infty}(4,8)$ code (quadrangle), and the $\mathcal{C}^{\infty}(5,10)$  (pentangle), each with the noisy Gaussian approximation algorithm. 
    Notice that the ordinate intercepts are $\varepsilon^{*}_{CRU}(3,6) = 0.8747$, $\varepsilon^{*}_{CRU}(4,8) = 0.8323$, and $\varepsilon^{*}_{CRU}(5,10) = 0.7910$, \cite[Table I]{ChungRU2001}.}
  \label{fig:gathresh}
\end{figure}

One might wonder whether there are sets of noise parameters $\alpha > 0$ and $\varepsilon > 0$ 
such that $s_{\ell} \to + \infty$.  Indeed there are, and there is a threshold phenomenon just like
Chung et al.\ showed for $\alpha = 0$ \cite{ChungRU2001}.
\begin{prop}
\label{prop:zero}
Final error probability $\eta^* = 0$ for LDPC ensembles decoded using the noisy Gaussian system defined in Section~\ref{sec:ex2}, 
for binary-input AWGN channels with noise level $\varepsilon < \varepsilon^*(\alpha)$.
\end{prop}
\begin{IEEEproof}
Substituting $s = +\infty$ into \eqref{eq:DEgauss} demonstrates that it is a stable fixed point.
It may further be verified that the dynamical system proceeds toward that fixed point if 
$\varepsilon < \varepsilon^*(\alpha)$.
\end{IEEEproof}
Unlike Section~\ref{sec:ex1} where the $\varepsilon^*(\eta,\alpha)$ thresholds
could be evaluated analytically, only numerical evaluations of these $\varepsilon^*(\alpha)$ 
thresholds are possible.  These are shown in Fig.~\ref{fig:gathresh} for three
regular LDPC ensembles with rate $1/2$, namely the (3,6) ensemble, the (4,8) ensemble, and the (5,10) ensemble.  
As can be observed, thresholds decrease smoothly as the decoder noise level increases.  Moreover, the ordering
of the codes remains the same for all levels of decoder noise depicted.  Code optimization remains to be done.

The basic reason for the disparity between Propositions \ref{prop:zero}
and \ref{prop:nonzero} is that here, the noise is bounded whereas the messages are unbounded.
Thus once the messages grow large, the noise has essentially no effect. 
To use a term from \cite{LentmaierTZC2005}, once the decoder reaches the \emph{breakout value}, noise 
cannot stop the decoder from achieving Shannon reliability. 

Perhaps a peak amplitude constraint on messages would provide a more realistic computation model, but the equivalent
of Proposition~\ref{prop:zero} may not hold.  Quantified data processing inequalities may
provide insight into what forms of noise and message constraints are truly limiting 
\cite{EvansS1999,Pippenger1988}.  

\section{Application: Reliable Memories Constructed from Unreliable Components}
\label{sec:Taylor}
In Section~\ref{sec:intro}, complexity and reliability were cast as the primary limitations
on practical decoding.  By considering the design of fault masking techniques for memory systems, 
a communication problem beyond Fig.~\ref{fig:fig1_rev}, both complexity and reliability may be 
explicitly constrained.  Indeed, the problem of constructing reliable information storage devices from 
unreliable components is central to fault-tolerant computing, and determining
the information storage capacity of such devices is a long-standing open problem \cite{ChilappagariVM2008}.
This problem is related to problems in distributed information storage \cite{DimakisR2008} and is
intimately tied to the performance of codes under faulty decoding.  The analysis techniques developed 
thus far may be used directly.

In particular, one may construct a memory architecture with noisy registers and 
a noisy LDPC correcting network.  At each time step,
the correcting network decodes the register contents and restores them.  
The correcting network prevents the codeword 
stored in the registers from wandering too far away.  Taylor and others have 
shown that there exist non-zero levels of component noisiness such that the 
LDPC-based construction achieves non-zero storage capacity \cite{Taylor1968,Kuznetsov1973,ChilappagariV2007}.  
Results as in Section~\ref{sec:ex1} may be used to precisely characterize storage capacity.

Before proceeding with an achievability result, requisite definitions and the problem 
statement are given \cite{Taylor1968}.
\begin{defin}
An \emph{elementary operation} is any Boolean function of two binary operands.
\end{defin}
\begin{defin}
A system is considered to be constructed from \emph{components}, which
are devices that either perform one elementary operation or store one bit.
\end{defin}
\begin{defin}
The \emph{complexity} $\chi$ of a system is the number of components within the system.
\end{defin}
\begin{defin}
A memory system that stores $k$ information bits is said to have an \emph{information
storage capability} of $k$.
\end{defin}
\begin{defin}
Consider a sequence of memories $\{M_i\}$, ordered according to their
information storage capability $i$ (bits).  The sequence $\{M_i\}$
is \emph{stable} if it satisfies the following:
\begin{enumerate}
  \item For any $k$, $M_k$ must have $2^k$ allowed inputs denoted $\{I_{k_i}\}$, $1\le i\le 2^k$.
  \item A class of states, $C(I_{k_i})$, is associated with each input $I_{k_i}$ of $M_k$.  The classes
	$C(I_{k_i})$ and $C(I_{k_j})$ must be disjoint for all $i\neq j$ and all $k$.
  \item The complexity of $M_k$, $\chi(M_k)$, must be bounded by $\theta k$, where \emph{redundancy} $\theta$ is fixed for all $k$.
  \item At $\ell = 0$, let one of the inputs from $\{I_{k_i}\}$ be stored in each memory $M_k$ in the sequence of memories $\{M_i\}$, 
	with no further inputs in times $\ell > 0$.  Let $I_{k_i}$ denote the particular input stored in memory $M_k$.
     Let $\lambda_{k_i}(T)$ denote the probability that the state of $M_k$ does not belong to $C(I_{k_i})$ at $\ell = T$ 
     and further let $P^{\max}_{k}(T) = \max_i \lambda_{k_i}(T)$.  Then for any $T>0$ and $\delta > 0$, there
     must exist a $k$ such that $P^{\max}_{k}(T) < \delta$.
\end{enumerate}
\end{defin}
The demarcation of classes of states is equivalent to demarcating decoding regions.
\begin{defin}
The \emph{storage capacity}, $\mathfrak{C}$, of memory is a number such that there exist stable memory sequences
for all memory redundancy values $\theta$ greater than $1/\mathfrak{C}$.
\end{defin}
Note that unlike channel capacity for the communication problem,
there is no informational definition of storage capacity that is known to go with 
the operational definition.

The basic problem then is to determine storage capacity,
which is a measure of the circuit complexity required
to achieve arbitrarily reliable information storage.  The circuit complexity
must be linear in blocklength, a property satisfied by systems with message-passing
correcting networks for LDPC codes.

Although Proposition~\ref{prop:nonzero} shows that Shannon reliability is not 
achievable for any noisy Gallager A decoder, the definition of stable 
information storage does not require this.  By only requiring maintenance within a decoding region,
the definition implies that either the contents of the memory may be read-out in coded form or equivalently that 
there is a noiseless output device that yields decoded information; call this noiseless output device the \emph{silver decoder}.

Consider the construction of a memory with noisy registers as storage
elements.  These registers are connected to a noisy Gallager A LDPC decoder (as described in Section~\ref{sec:ex1}), 
which takes the register values as inputs and stores its computational results back into 
the registers.  To find the storage capacity of this construction, first compute the
complexity (presupposing that the construction will yield a stable sequence of memories).  

The Gallager A check node operation 
is a $(\tr-1)$-input XOR gate, which may be constructed from $\tr - 2$ two-input XOR gates.  
A variable node determines whether its $\tl - 1$ inputs are all the same and then compares
to the original received value.  Let $D_{\tl}$ denote the complexity of this logic.
The output of the comparison to the original received value is the value of the consensus 
view.  One construction to implement the consensus logic is to OR together the outputs of a
$(\tl - 1)$-input AND gate and a $(\tl - 1)$-input AND gate with inverted inputs.  This is then XORed with
the stored value.  Such a circuit can be implemented with $2(\tl - 2) + 2$ components, so $D_{\tl} = 2\tl - 2$. 
The storage is carried out in $n$ registers.  The total complexity of the memory
$M_k$, $\chi(M_k)_{\mathcal{C}^n(\tl,\tr)}$, is
\[
\chi(M_k)_{\mathcal{C}^n(\tl,\tr)} = n(1 + 2\tl - 2 + \tl(\tr-2)) = n(\tl\tr - 1) \mbox{.}
\]

The information storage capability is $n$ times the rate of the code, $R$.  The complexity
of an irredundant memory with the same storage capability is $\chi_{\mbox{irr}_n} = Rn$.
Hence, the redundancy is
\[
\frac{\chi(M_k)_{\mathcal{C}^n(\tl,\tr)}}{\chi_{\mbox{irr}_n}} = \frac{n(\tl\tr - 1)}{Rn} \le \frac{(\tl\tr - 1)}{1 - \tl/\tr}
\]
which is a constant.  By \cite[Lemma 3.22]{RichardsonU2008}, the inequality 
almost holds with equality with high probability for large $n$.  For the 
$(3,6)$ regular LDPC code, the redundancy value is $34$, so $\mathfrak{C} = 1/34$,
if the construction does in fact yield stable memories.

The conditions under which the memory is stable depends on the silver decoder.
Since silver decoder complexity does not enter, maximum likelihood 
should be used.  The Gallager lower bound to the ML decoding threshold for the $(3,6)$ regular LDPC 
code is $\varepsilon^*_{GLB} = 0.0914755$ \cite[Table II]{Montanari2005}.  
Recall from Fig.~\ref{fig:region_to_use_decoder} that the decoding threshold for 
Gallager A decoding is $\varepsilon^*_{BRU} = 0.0394636562$.

If the probability of bit error for the correcting network in the memory stays
within the decoding threshold of the silver decoder, then stability follows.
Thus the question reduces to determining the sets of component noisiness levels 
$(\alpha,\varepsilon)$ for which the decoding circuit achieves 
$(\eta = \varepsilon^*_{ML})$-reliability.  

\begin{figure}
  \centering
  \includegraphics[width=3in]{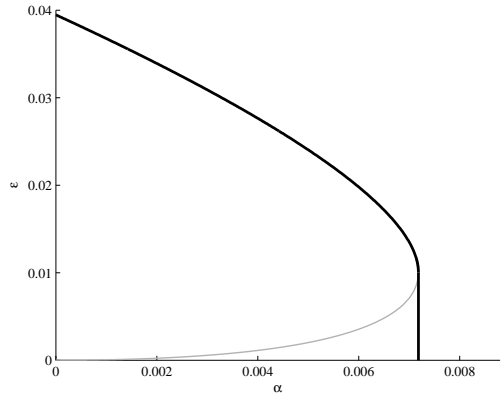}
  \caption{For a memory system constructed with noisy registers and a $(3,6)$ LDPC Gallager A correcting network, 
	the region $\mathfrak{R}$ (delimited by black line) comprises the ``region to use decoder'' and its hypograph.}
  \label{fig:noisyreg_r}
\end{figure}

Consider a memory system where bits are stored in registers
with probability $\alpha_r$ of flipping at each time step.  An LDPC codeword
is stored in these registers; the probability of incorrect storage at the first 
time step is $\varepsilon$.  At each iteration, the variable node value
from the correcting network is placed in the register.  This stored value 
is used in the subsequent Gallager A variable node computation rather than
a received value from the input pins.  Suppose that the component noise values 
in the correcting network may be parameterized as in Section~\ref{sec:ex1}.
Then a slight modification of the analysis in Section~\ref{sec:ex1} yields 
a density evolution equation
\[
s_{\ell + 1} = \varepsilon_2 - \varepsilon_2 q_{\alpha}^{+}(s_{\ell}) + (1-\varepsilon_2) q_{\alpha}^{-}(s_{\ell}) \mbox{,}
\]
where $\varepsilon_2 = s_{\ell}(1-\alpha_r) + \alpha_r(1-s_{\ell})$.
There is a ``region to use decoder'' for this system, just as in Section~\ref{sec:ex1}.
If $\alpha_r = \alpha$, this region is shown in Fig.~\ref{fig:noisyreg_r},
and is slightly smaller than the region in Fig.~\ref{fig:region_to_use_decoder}.
Denote this region and its hypograph as $\mathfrak{R}$.  It follows
that $(\eta = \varepsilon^*_{BRU})$-reliability is achieved for $\mathfrak{R}$.
Since $\varepsilon^*_{BRU}$-reliability 
is achievable, $\varepsilon^*_{GLB}$-reliability is achievable by 
monotonicity.  Thus the construction yields stable memories.
\begin{prop}
Let $\mathfrak{R}$ be the set of memory component noise parameters $(\alpha,\varepsilon)$ 
within the region to use decoder or its hypograph corresponding to a system with a 
Gallager A correcting network for the $(3,6)$ LDPC code, depicted in Fig.~\ref{fig:noisyreg_r}. 
Then a sequence of memories constructed from $\mathfrak{R}$-components have a storage capacity 
lower bounded as $\mathfrak{C} \ge 1/34$.
\end{prop}
This may be directly generalized for any choice of code ensemble as follows.
\begin{thm}
Let $\mathfrak{R}$ be the (computable) set of memory component noise parameters $(\alpha,\varepsilon)$ 
within the region to use decoder or its hypograph corresponding to a system with a 
Gallager A correcting network for the $(\lambda,\rho)$ LDPC code. 
Then a sequence of memories constructed from $\mathfrak{R}$-components have a storage capacity 
lower bounded as
\[
\mathfrak{C} \ge \frac{1 - \lambda^{\prime}(1)/\rho^{\prime}(1)}{\lambda^{\prime}(1)\rho^{\prime}(1) - 1} \mbox{.}
\]
The bound reduces to $({1 - \tl/\tr})/({\tl\tr - 1})$ for regular codes.
\end{thm}
This theorem gives a precise achievability result that bounds storage capacity.
It also implies a code ensemble optimization problem similar to the one in Section~\ref{sec:optim}.  
The question of an optimal architecture for memory systems however remains open.

\section{Conclusions}
\label{sec:conc}
Loeliger et al.~\cite{LoeligerLHT2001} had observed that decoders are robust to nonidealities and
noise in physical implementations, however they had noted that 
``the quantitative analysis of these effects is a challenging 
theoretical problem.''  This work has taken steps to address 
this challenge by characterizing robustness to decoder noise.

The extension of the density evolution method to the case of
faulty decoders allows a simplified means of asymptotic
performance characterization.  Results from this method show
that in certain cases Shannon reliability
is not achievable (Proposition~\ref{prop:nonzero}),
whereas in other cases it
is achievable (Proposition~\ref{prop:zero}).  In either case,
however, the degradation of a suitably defined decoding threshold
is smooth with increasing decoder noise, whether in circuit nodes or circuit wires.
Due to this smoothness, codes optimized for fault-free decoders do work well
with faulty decoders, however optimization of codes for systems with faulty decoders remains to be studied.

No attempt was made to apply fault masking methods to develop decoding algorithms with 
improved performance in the presence of noise.  One approach might be to use coding within the decoder
so as to reduce the values of $\alpha$.  Of course, the within-decoder code
would need to be decoded.  There are also more direct circuit-oriented techniques
that may be applied \cite{Tryon1962,GaoQF2005}.  Following the concept of
concatenated codes, concatenated decoders may also be promising.  The basic idea
of using a first (noiseless) decoder to correct many errors and
then a second (noiseless) decoder to clean things up was already present in
\cite{LubyMSS2001}, but it may be extended to the faulty decoder setting.

Reducing power consumption in decoder circuits has been an active area of research
\cite{NawabOCWL1997,SinhaWC2002,BhardwajMC2001,HegdeS2000,WangS2003,MaseraMPVZ2002,MansourS2006,SahaiG2008},
however power reduction often has the effect of increasing noise in the decoder \cite{MaheshwariBT2004}.
The tradeoff developed between the quality of the communication channel and
the quality of the decoder may provide guidelines for allocating resources
in communication system design.

Analysis of other
decoding algorithms with other error models will presumably
yield results similar to those obtained here.  For greater generality, one might move beyond
simple LDPC codes and consider arbitrary codes decoded with very general
iterative decoding circuits \cite{SahaiG2008} with suitable error models.  
An even more general model of computation such as a Turing machine or beyond 
\cite{Copeland2002} does not seem to have an obvious, appropriate error model.

Even just a bit of imagination provides numerous models of channel noise and
circuit faults that may be investigated in the future to provide further
insights into the fundamental limits of noisy communication and computing.

\section*{Acknowledgment}
I thank R\"{u}diger L.\ Urbanke and \.{I}.\ Emre Telatar for several enlightening discussions, for encouragement,
and for hosting my visit at EPFL.  I also thank the anonymous reviewers, Sanjoy K.\ Mitter,
G.\ David Forney, and Vivek K Goyal for assistance in improving the paper.  Thanks also
to Shashi Kiran Chilappagari for telling me about his work.

\appendices
\section{Proof of Theorem~\ref{thm:symmetry}}
\label{app:symmetry}
Let ${\bf x} \in \mathcal{C}^n$ be a codeword and let $\bf{Y}$ denote the 
corresponding channel output $\bf{Y} = \bf{x}\bf{Z}$ (where the notation means 
pointwise multiplication on length $n$ vectors).  Note that $\bf{Z}$ is
equal to the channel output observation when $\bf{x}$ is all-one.  The goal 
is to show that messages sent during the decoding process for cases when the
received codeword is either $\bf{x}\bf{Z}$ or $\bf{x}$ correspond.

Let $\dot{n}_i$ be an arbitrary variable node and let $\dot{n}_j$ be one
of its neighboring check nodes.  Let $\nu_{ij}^{(\ell)}(\bf{y})$ and $\mu_{ij}^{(\ell)}(\bf{y})$ 
denote the variable-to-check
message from $\dot{n}_i$ to $\dot{n}_j$ at the respective terminals in iteration $\ell$, 
assuming received value $\bf{y}$.  Similarly, let $\nu_{ji}^{(\ell)}(\bf{y})$ and $\mu_{ji}^{(\ell)}(\bf{y})$ 
be the check-to-variable message from $\dot{n}_j$ to $\dot{n}_i$ at the respective terminal in 
iteration $\ell$ assuming received value $\bf{y}$.

By Definition~\ref{def:def1}, the channel is memoryless binary-input output-symmetric and
it may be modeled multiplicatively as
\begin{equation}
\label{eq:multiplicatively}
Y_t = x_t Z_t \mbox{,}
\end{equation}
where $\{Z_t\}$ is a sequence of i.i.d.\ random variables and $t$ is the channel usage time.  
The validity of the multiplicative model is shown in \cite[p.~605]{RichardsonU2001} and 
\cite[p.~184]{RichardsonU2008}.

The proof proceeds by induction and so the base case is established first.
By the multiplicative model \eqref{eq:multiplicatively}, $\nu_{ij}^{(0)}({\bf y}) = \nu_{ij}^{(0)}({\bf x z})$.
Recalling that $x_i \in \{\pm 1\}$, by the variable node symmetry condition (Definition $3$) 
which includes computation noise $u_{\dot{n}_i}^{(0)}$, 
it follows that $\nu_{ij}^{(0)}({\bf y}) = \nu_{ij}^{(0)}({\bf x z})= x_i\nu_{ij}^{(0)}({\bf z})$.  

Now take the wire noise $w_{ij}^{(0)}$ on the message from $\dot{n}_i$ to $\dot{n}_j$ into account.  It is
symmetric (Definition~\ref{def:def4}) and so $\nu_{ij}^{(0)}({\bf y}) =  x_i\nu_{ij}^{(0)}({\bf z})$ implies a similar
property for $\mu_{ij}^{(0)}$.  In particular, 
\begin{align}
\label{eq:wirenoisesym}
\mu_{ij}^{(0)}({\bf y}) &= \Xi(\nu_{ij}^{(0)}({\bf y}), w_{ij}^{(0)}) \\ \notag
&= \Xi(x_i\nu_{ij}^{(0)}({\bf z}),w_{ij}^{(0)})  \\ \notag
&= x_i \Xi(\nu_{ij}^{(0)}({\bf z}),x_i w_{ij}^{(0)}) 
\end{align}
where the last step follows because $x_i \in \{\pm 1\}$ and so it can be taken outside of $\Xi$
by Definition~\ref{def:def4}, when it is put back in for the wire noise. Now since $x_i \in \{\pm 1\}$ and 
since the wire noise is symmetric about $0$ by Definition~\ref{def:def4},  
$x_i \Xi(\nu_{ij}^{(0)}({\bf z}),x_i w_{ij}^{(0)})$ will correspond to
$x_i\mu_{ij}^{(0)}({\bf z})$, in the sense that error event probabilities will be identical.

Assume that $\mu_{ij}^{(\ell)}({\bf y})$ corresponds to $x_i\mu_{ij}^{(\ell)}({\bf z})$ for all $(i,j)$ pairs
and some $\ell \ge 0$ as the inductive assumption.  Let $\mathcal{N}_{\dot{n}_j}$ be the set of all variable nodes that 
are connected to check node $\dot{n}_j$.  Since $\bf{x}$ is a codeword, it satisfies the parity checks,
and so $\prod_{k\in \mathcal{N}_{\dot{n}_j}} = 1$.  Then from the check node symmetry 
condition (Definition~\ref{def:def2}), $\nu_{ji}^{(\ell + 1)}({\bf y})$ corresponds to $x_i\nu_{ji}^{(\ell + 1)}({\bf z})$.
Further, by the wire noise symmetry condition (Definition~\ref{def:def4}) and the same argument as for the base case,  
$\mu_{ji}^{(\ell + 1)}({\bf y})$ corresponds to $x_i\mu_{ji}^{(\ell + 1)}({\bf z})$.
By invoking the variable node symmetry condition (Definition~\ref{def:def3}) again, 
it follows that $\nu_{ij}^{(\ell + 1)}({\bf y})$ corresponds to $x_i\nu_{ij}^{(\ell + 1)}({\bf z})$
for all $(i,j)$ pairs.  

Thus by induction, all messages to and from 
variable node $\dot{n}_i$ when $\bf{y}$ is received correspond to
the product of $x_i$ and the corresponding message when $\bf{z}$ is received.

Both decoders proceed in correspondence and commit exactly the same number of errors.

\subsubsection{Worst-Case Noise}
\label{app:worst}
The same result with the same basic proof also holds when the wire noise operation $\Xi$ is symmetric but 
$w$ is not symmetric stochastic, but is instead worst-case.
The only essential modification is in \eqref{eq:wirenoisesym} and the related part of the induction
step.  Since wire noise is dependent on $x_i$, it can be written as $x_i w$.  Thus,
\begin{align*}
\mu_{ij}^{(0)}({\bf y}) &= \Xi(\nu_{ij}^{(0)}({\bf y}), x_i w_{ij}^{(0)}) \\ \notag
&= \Xi(x_i\nu_{ij}^{(0)}({\bf z}),x_i w_{ij}^{(0)})  \\ \notag
&\stackrel{(a)}{=} x_i \Xi(\nu_{ij}^{(0)}({\bf z}),w_{ij}^{(0)}) \\ \notag
&= x_i\mu_{ij}^{(0)}({\bf z})
\end{align*}
where step (a) follows because $x_i \in \{\pm 1\}$ and so it can be taken outside of $\Xi$
by the symmetry property of $\Xi$.  Thus the two decoders will proceed in exact
one-to-one correspondence, not just in probabilistic correspondence.

\section{Proof of Theorem~\ref{thm:conc}}
\label{app:concentrate}
Prior to giving the proof of Theorem~\ref{thm:conc}, 
a review of some definitions from probability theory \cite{Williams1991} and 
the Hoeffding-Azuma inequality are provided.

Consider a measurable space $(\Omega,\mathcal{F})$ consisting of a sample space
$\Omega$ and a $\sigma$-algebra $\mathcal{F}$ of subsets of $\Omega$ that contains
the whole space and is closed under complementation and countable unions.  A random
variable is an $\mathcal{F}$-measurable function on $\Omega$.  If there is a
collection $(Z_{\gamma} | \gamma \in C)$ of random variables $Z_{\gamma}: \Omega \to \mathbb{R}$, then
\[
\mathcal{Z} = \sigma(Z_{\gamma} | \gamma \in C)
\]
is defined to be the smallest $\sigma$-algebra $\mathcal{Z}$ on $\Omega$ such that
each map $(Z_{\gamma}|\gamma \in C)$ is $\mathcal{Z}$-measurable.  

\begin{defin}[Filtration]
Let $\{\mathcal{F}_i\}$ be a sequence of $\sigma$-algebras with respect to the same 
sample space $\Omega$.  These $\mathcal{F}_i$ are said to form a \emph{filtration}
if $\mathcal{F}_0 \subseteq \mathcal{F}_1 \subseteq \cdots$ are ordered
by refinement in the sense that each subset of $\Omega$ in $\mathcal{F}_i$ is
also in $\mathcal{F}_j$ for $i \le j$.  Also $\mathcal{F}_0 = \{\emptyset,\Omega  \}$.
\end{defin} 
Usually, $\{\mathcal{F}_i\}$ is the \emph{natural filtration} $\mathcal{F}_i = \sigma(Z_0,Z_1,\ldots,Z_i)$
of some sequence of random variables $(Z_0,Z_1,\ldots)$, and then the knowledge about $\omega$
known at step $i$ consists of the values $Z_0(\omega),Z_1(\omega),\ldots,Z_i(\omega)$.

For a probability triple $(\Omega,\mathcal{F},\mathbb{P})$, 
a version of the conditional expectation of a random variable $Z$ given a $\sigma$-algebra 
$\mathcal{F}$ is a random variable denoted $\E{Z|\mathcal{F}}$.  Two versions of
conditional expectation agree almost surely, but measure zero departures are not considered subsequently;
one version is fixed as canonical.
Conditional expectation given a measurable event $\mathfrak{E}$ is denoted $\E{Z|\sigma(\mathfrak{E})}$ and
conditional expectation given a random variable $W$ is denoted $\E{Z|\sigma(W)}$.  

\begin{defin}[Martingale]
Let $\mathcal{F}_0 \subseteq \mathcal{F}_1 \subseteq \cdots$ be a filtration 
on $\Omega$ and let $Z_0,Z_1,\ldots$ be a sequence of random variables on 
$\Omega$ such that $Z_i$ is $\mathcal{F}_i$-measurable.  Then $Z_0,Z_1,\ldots$
is a \emph{martingale} with respect to the filtration 
$\mathcal{F}_0 \subseteq \mathcal{F}_1 \subseteq \cdots$ if 
$\E{Z_i|\mathcal{F}_{i-1}} = Z_{i-1}$.
\end{defin}
A generic way to construct a martingale is Doob's construction.
\begin{defin}[Doob Martingale]
Let $\mathcal{F}_0 \subseteq \mathcal{F}_1 \subseteq \cdots$ be a filtration 
on $\Omega$ and let $Z$ be a random variable on $\Omega$.  Then the sequence
of random variables $Z_0,Z_1,\ldots$ such that $Z_i = \E{Z|\mathcal{F}_i}$
is a Doob martingale.
\end{defin}

\begin{lemma}[Hoeffding-Azuma Inequality \cite{Azuma1967,Hoeffding1963,RichardsonU2001}]
Let $Z_0,Z_1,\ldots$ be a martingale with respect to the filtration 
$\mathcal{F}_0 \subseteq \mathcal{F}_1 \subseteq \cdots$ such that
for each $i > 0$, the following bounded difference condition is satisfied
\[
|Z_i - Z_{i-1}| \le \alpha_i \mbox{, }\alpha_i \in [0,\infty)\mbox{.}
\]
Then for all $n > 0$ and any $\xi > 0$,
\[
\Pr\left[|Z_n - Z_0| \ge \xi\right] \le 2\exp\left( -\frac{\xi^2}{2\sum_{k=1}^n \alpha_k^2} \right) \mbox{.}
\]
\end{lemma}

Now to the proof of Theorem~\ref{thm:conc}; as noted before, it is an extension 
of \cite[Theorem 2]{RichardsonU2001} or \cite[Theorem 4.94]{RichardsonU2008}.
The basic idea is to construct a Doob martingale about the 
object of interest by revealing various randomly determined aspects in a 
filtration-refining manner.  The first set of steps is used to reveal
which code was chosen from the ensemble of codes; the $n\tl$ edges 
in the bipartite graph are ordered in some arbitrary manner and exposed
one by one.  Then the $n$ channel noise realizations are revealed.  At
this point the exact graph and the exact channel noise realizations encountered
have been revealed.  Now the decoder noise realizations must be revealed.
There are $n$ variable nodes, so the computation noise in each of them is 
revealed one by one.  There are $n\tl$ edges over which variable-to-check 
communication noise is manifested.  Then there are $n\tl/\tr$ check nodes
with computation noise, and finally 
there are $n\tl$ check-to-variable communication noises for one 
iteration of the algorithm.  The decoder noise realizations
are revealed for each iteration.  At the beginning of the revelation process,
the average (over choice of code, channel noise realization, and decoder
noise realization) is known; after the $m = (\tl + 2\ell\tl + 1 + \ell + \ell\tl/\tr)n$ 
revelation steps, the exact system used is known.  

Recall that $Z$ denotes the number of incorrect values held at the end of the $\ell$th 
iteration for a particular $(g,y,w,u) \in \Omega$.  Since $g$ is a graph in the 
set of labeled bipartite factor graphs with variable node degree 
$\tl$ and check node degree $\tr$, $\mathcal{G}^n(\tl,\tr)$; $y$ is a particular input to the decoder, $y \in \mathcal{Y}^n$;
$w$ is a particular realization of the message-passing noise, $w \in \mathcal{M}^{2\ell\tl n}$; and $u$ is a particular 
realization of the local computation noise, $u \in \mathcal{U}^{(\ell + \ell\tl/\tr)n}$,
the sample space is 
$\Omega = \mathcal{G}^n(\tl,\tr) \times \mathcal{Y}^n \times \mathcal{M}^{2\ell\tl n} \times \mathcal{U}^{(\ell + \ell\tl/\tr)n}$.  

In order to define random variables, first define the following exposure procedure.
Suppose realizations of random quantities are exposed sequentially.  
First expose the $\tl n$ edges of the graph one at
a time.  At step $i \le \tl n$ expose the particular check node
socket which is connected to the $i$th variable node socket.
Next, in the following $n$ steps, expose the received
values $y_i$ one at a time.  Finally in the remaining $(2\tl + 1 + \tl/\tr) \ell n$ steps, 
expose the decoder noise values $u_i$ and $w_i$ that were encountered 
in all iterations up to iteration $\ell$.

Let $\equiv_i$, $0 \le i \le m$, be a sequence of equivalence relations on 
the sample space $\Omega$ ordered by refinement.
Refinement means that $(g',y',w',u') \equiv_i (g'',y'',w'',u'')$ implies $(g',y',w',u') \equiv_{i-1} (g'',y'',w'',u'')$.
The equivalence relations define equivalence classes such that $(g',y',w',u') \equiv_i (g'',y'',w'',u'')$ if
and only if the realizations of random quantities revealed in the first $i$ steps for both
pairs is the same.  

Now, define a sequence of random variables $Z_0, Z_1, \ldots, Z_m$.  Let the random variable 
$Z_0$ be $Z_0 = \E{Z}$, where the expectation is over the code choice, channel noise , and decoder
noise.  The remaining random variables $Z_i$ are constructed as conditional expectations
given the measurable equivalence events $(g',y',w',u') \equiv_i (g,y,w,u)$: 
\[
Z_i(g,y,w,u) = \E{Z(g',y',w',u') | \sigma((g',y',w',u') \equiv_i (g,y,w,u))} \mbox{.}
\]
Note that $Z_m = Z$ and that by construction $Z_0, Z_1, \ldots, Z_m$ is a Doob martingale.
The filtration is understood to be the natural filtration of the random variables $Z_0,Z_1,\ldots,Z_m$.

To use the Hoeffding-Azuma inequality to give bounds on
\[
\Pr\left[|Z - \E{Z}| > n \tl \epsilon /2\right] = \Pr\left[|Z_m - Z_0| > n \tl \epsilon /2\right] \mbox{,}
\]
bounded difference conditions 
\[
|Z_{i+1}(g,y,w,u) - Z_{i}(g,y,w,u)| \le \alpha_i \mbox{,  } i = 0,\ldots,m-1
\]
need to be proved for suitable constants $\alpha_i$ that may depend 
on $\tl$, $\tr$, and $\ell$.

For the steps where bipartite graph edges are exposed, it was shown in 
\cite[p.\ 614]{RichardsonU2001} that
\[
|Z_{i+1}(g,y,w,u) - Z_{i}(g,y,w,u)| \le 8(\tl\tr)^{\ell} \mbox{, } 0 \le i < n \tl \mbox{.}
\]
It was further shown in \cite[p.\ 615]{RichardsonU2001} that for the steps when the channel outputs 
are revealed that 
\begin{equation}
\label{eq:boundeddifference_y}
|Z_{i+1}(g,y,w,u) - Z_{i}(g,y,w,u)| \le 2(\tl\tr)^{\ell} \mbox{, } n \tl \le i < n(1 + \tl) \mbox{.}
\end{equation}
It remains to show that the inequality is also fulfilled for steps when
decoder noise realizations are revealed.  The bounding procedure is nearly identical 
to that which yields \eqref{eq:boundeddifference_y}.
When a node noise realization $u$
is revealed, clearly only something whose directed neighborhood includes the node
at which the noise $u$ causes perturbations can be affected.  Similarly, 
when an edge noise realization $w$ is revealed,
only something whose directed neighborhood includes the edge on which the noise $w$ causes perturbations
can be affected.  In \cite[p.\ 603]{RichardsonU2001}, it is shown that the size of the directed neighborhood
of depth $2\ell$ of the node $\dot{n}(u)$ associated with noise $u$ is bounded as 
$|\mathcal{N}_{\dot{n}(u)}^{2\ell}| \le 2(\tl\tr)^{\ell}$ and similarly the size
of the directed neighborhood of length $2\ell$ of the edge $\vec{e}(w)$ associated 
with noise $w$ is bounded as $|\mathcal{N}_{\vec{e}(w)}^{2\ell}| \le 2(\tl\tr)^{\ell}$.  
Since the maximum depth that can be affected by a noise perturbation is $2\ell$, 
a weak uniform bound for the remaining exposure steps is
\[
|Z_{i+1}(g,y,w,u) - Z_{i}(g,y,w,u)| \le 2(\tl\tr)^{\ell} \mbox{, } n(1 + \tl) \tl \le i < m \mbox{.}
\]
Since bounded difference constants $\alpha_i$ have been provided for all $i$, 
the theorem follows from application of the Hoeffding-Azuma inequality to 
the martingale.  

One may compute a particular value of $\beta$ to use as follows.  The bounded difference sum is
\begin{align*}
\sum_{k=1}^m \alpha_k^2 &= 64n\tl (\tl\tr)^{2\ell} + 4n(\tl\tr)^{2\ell} + 4[2\ell\tl n + n\ell + n\ell\tl /\tr](\tl\tr)^{2\ell} \\ \notag
&= n\left\{64 \tl + 4 + 8\tl\ell + \ell + \tfrac{\tl\ell}{\tr}\right\} \tl^{2\ell}\tr^{2\ell}
\end{align*}
Setting constants in the theorem and in the Hoeffding-Azuma inequality equal yields
\begin{align*}
\tfrac{1}{\beta} &= 512 \tl^{2\ell - 1}\tr^{2\ell} + 32 \tl^{2\ell - 2}\tr^{2\ell} + 64\ell \tl^{2\ell - 1}\tr^{2\ell} + 8\ell\tl^{2\ell-1}\tr^{2\ell - 1} + 8\ell \tl^{2\ell-2}\tr^{2\ell} \\ \notag
&\le (544 + 80\ell)\tl^{2\ell-1}\tr^{2\ell} 
\end{align*}
Thus $\tfrac{1}{\beta}$ can be taken as $(544 + 80\ell)\tl^{2\ell-1}\tr^{2\ell}$. 

\section{An Analytical Expression}
\label{app:analexpr}

An analytical expression for $\varepsilon^*(\eta = 1/10, \alpha = 5\times 10^{-3})$
is
\[
\tfrac{1}{2} \left(1 - \sqrt{1 + 4c_7}\right) \mbox{,}
\]
where $c_7$ is the second root of the polynomial in $\check{\varepsilon}$
\[
c_1 + c_2\check{\varepsilon} + c_3\check{\varepsilon}^2 + c_4\check{\varepsilon}^3 + c_5\check{\varepsilon}^4 + c_6\check{\varepsilon}^5 \mbox{,}
\]
and constants $(c_1,\ldots,c_6)$ are defined as follows.
\begin{align*}
c_1 &= 36 \alpha^2 - 360 \alpha^3 + 1860\alpha^4 - 6240 \alpha^5 + 14752 \alpha^6 - 25344 \alpha^7 + 31680 \alpha^8 \\ \notag
&\qquad  - 28160 \alpha^9 + 16896 \alpha^{10} - 6144 \alpha^{11} + 1024 \alpha^{12} \\ \notag
    &= \frac{3424572914129280658801}{4000000000000000000000000}
\end{align*}
\begin{align*}
c_2 &= 1 - 72 \alpha + 1080 \alpha^2 - 8160 \alpha^3 + 38640 \alpha^4 - 125952 \alpha^5 + 295424 \alpha^6 - 506880 \alpha^7 \\ \notag
&\qquad + 633600 \alpha^8 - 563200 \alpha^9 + 337920 \alpha^{10} - 122880 \alpha^{11} + 20480 \alpha^{12} \\ \notag
	&= \frac{133200752195329280658801}{200000000000000000000000}
\end{align*}
\begin{align*}
c_3 &= 32 - 864 \alpha + 10080 \alpha^2 - 69120 \alpha^3 + 314880 \alpha^4 - 1012224 \alpha^5 + 2364928 \alpha^6 - 4055040 \alpha^7\\ \notag
	&\qquad  + 5068800 \alpha^8 - 4505600 \alpha^9 + 2703360 \alpha^{10} - 983040 \alpha^{11} + 163840 \alpha^{12} \\ \notag
	&= \frac{698088841835929280658801}{25000000000000000000000}
\end{align*}
\begin{align*}
c_4 &= 160 - 3840 \alpha + 42240 \alpha^2 - 281600 \alpha^3 + 1267200 \alpha^4 - 4055040 \alpha^5 + 9461760 \alpha^6 - 16220160 \alpha^7 \\ \notag
	&\qquad  + 20275200 \alpha^8 - 18022400 \alpha^9 + 10813440 \alpha^{10} - 3932160 \alpha^{11} + 655360 \alpha^{12} \\ \notag
	&= \frac{886384871716129280658801}{6250000000000000000000}
\end{align*}
\begin{align*}
c_5 &= 320 - 7680 \alpha + 84480 \alpha^2 - 563200 \alpha^3 + 2534400 \alpha^4 - 8110080 \alpha^5 + 18923520 \alpha^6 - 32440320 \alpha^7 \\ \notag
	&\qquad  + 40550400 \alpha^8 - 36044800 \alpha^9 + 21626880 \alpha^{10} - 7864320 \alpha^{11} + 1310720 \alpha^{12} \\ \notag
	&= \frac{886384871716129280658801}{3125000000000000000000}
\end{align*}
\begin{align*}
c_6 &= 256 - 6144 \alpha + 67584 \alpha^2 - 450560 \alpha^3 + 2027520 \alpha^4 - 6488064 \alpha^5 + 15138816 \alpha^6 - 25952256 \alpha^7\\ \notag
	&\qquad  + 32440320 \alpha^8 - 28835840 \alpha^9 + 17301504 \alpha^{10} - 6291456 \alpha^{11} + 1048576 \alpha^{12} \\ \notag
	&= \frac{886384871716129280658801}{3906250000000000000000}
\end{align*}

As given in Table~\ref{tab:performance}, the numerical value of 
$\varepsilon^*(\eta = 1/10, \alpha = 5\times 10^{-3})$ is $0.0266099758$.

Similarly complicated analytical expressions are available for the other entries of 
Table~\ref{tab:performance} and the values used to create 
Figs.~\ref{fig:fixed_alpha_plot}, \ref{fig:region_to_use_decoder}, and \ref{fig:contourplot}.

\bibliographystyle{IEEEtran} 
\bibliography{abrv,conf_abrv,lrv_lib}

\end{document}